\documentclass[pdflatex,sn-mathphys]{sn-jnl}

\jyear{2022}%

\theoremstyle{thmstyleone}%
\usepackage{bm}
\theoremstyle{thmstyletwo}%

\theoremstyle{thmstylethree}%

\begin{document}

\title[Non-Fourier heat transport in nanosystems]{Non-Fourier heat transport in nanosystems}


\author[1,2]{\fnm{Giuliano} \sur{Benenti}}\email{giuliano.benenti@uninsubria.it}
\author[3]{\fnm{Davide} \sur{Donadio}}\email{ddonadio@ucdavis.edu}
\author*[4]{\fnm{Stefano} \sur{Lepri}}\email{stefano.lepri@isc.cnr.it}
\author[5,4]{\fnm{Roberto} \sur{Livi}}\email{roberto.livi@unifi.it}

\affil[1]{\orgdiv{Center for Nonlinear and Complex Systems, Dipartimento di
Scienza e Alta Tecnologia}, \orgname{Universit\`a degli Studi dell'Insubria}, 
\orgaddress{\street{via Valleggio 11}, \city{Como}, \postcode{22100} \state{}, \country{Italy}}}

\affil[2]{\orgdiv{Istituto Nazionale di Fisica Nucleare, Sezione di Milano}, 
\orgaddress{\street{via Celoria 16}, \city{Milano}, \postcode{20133} \state{}, \country{Italy}}}

\affil[3]{\orgdiv{Department of Chemistry}, \orgname{University of California Davis}, \orgaddress{\street{One Shields Ave.}, \city{Davis}, \postcode{95616}, \state{CA}, \country{USA}}}

\affil[4]{\orgdiv{Istituto dei Sistemi Complessi}, \orgname{Consiglio Nazionale delle Ricerche}, \orgaddress{\street{via Madonna del piano 10}, \city{Sesto fiorentino}, \postcode{I-50019} \state{}, \country{Italy}}}

\affil[5]{\orgdiv{Dipartimento di Fisica e Astronomia}, \orgname{Universita di Firenze}, \orgaddress{\street{Via G. Sansone 1}, \city{Sesto fiorentino}, \postcode{I-50019} \state{}, \country{Italy}}}


\abstract{Energy transfer in small nanosized systems can be very different
from that in their macroscopic counterparts due to reduced dimensionality, interaction with surfaces, disorder, and large fluctuations. Those ingredients may
induce non-diffusive heat transfer that requires to be taken into account
on small scales. We provide an overview of the
recent advances in this field from the points of view of nonequilibrium
statistical mechanics and atomistic simulations.
We summarize the underlying basic properties leading to violations
of the standard diffusive picture of heat transport and its universal
features, with some historical perspective. 
We complete this scenario by illustrating also the effects
of long-range interaction and integrability on non-diffusive
transport. Then we discuss how all of these features can be exploited
for thermal management, rectification and to improve 
the efficiency of energy conversion.
We conclude with a review on recent achievements in 
atomistic simulations of 
anomalous heat transport in single polymers, nanotubes and 
two-dimensional materials. A short account of the existing experimental
literature is also given.}

\keywords{Nonequilibrium statistical mechanics, Anomalous transport, Thermal conversion, Coupled Transport, Atomistic simulations}



\maketitle

\section{Introduction}
\label{sec1}

Energy transfer in nonlinear systems occurs in many physical
contexts, ranging from condensed matter to optics. Besides
its basic interest,   
understanding the principles of vibrational energy transport at the nanoscale
is of importance
to eventually improve our capability to manage thermal 
transport on small-scale devices \cite{volz2016nanophononics}.
In the case of phononic systems, where heat is mostly transported 
by lattice vibrations, this calls
for a deeper understanding of the properties of strongly anharmonic
and/or disordered crystals and artificial materials.
Nonlinear effects are essential in many respects: in the 
first place, they determine thermal
transport properties. When reducing the system size towards  the 
micro and nano-scale a series of novel effects appear. 
Temperature gradients can become very large, leading to non-linear 
response. Also, the role of thermal contact and interfaces 
may not be negligible, and in general heat-current fluctuations
may be relevant and one has to treat them within a suitable thermodynamic approach for small systems. Particularly 
dramatic effects occur in reduced dimensions, 
where nonlinear interactions of energy fluctuations lead to anomalous 
conductivity.

In this article, we will review the main features of non-Fourier (superdiffusive)
heat conduction in low-dimensional 
many-body systems. Starting from simple
theoretical models (anharmonic chains)
we will argue that this phenomenon 
displays remarkable universal properties.
As it is known, 
the idea of universality is very relevant
in statistical physics: it tells us that some quantitative
relations hold, independently of the microscopic details. The concept was
born out of the theory of critical phenomena: for instance, any ferromagnetic transition will have the same dependence of the
magnetization on temperature if the underlying Hamiltonian has the same
symmetry.

In the context of nonequilibrium processes (like heat transport), this is even more striking. For instance, if we have a nearly one-dimensional system with some symmetries leading to momentum, energy, and density conservation, we expect superdiffusive transport with given exponents, independently of the actual microscopic forces.
As a concrete example, the simple Fermi-Pasta-Ulam-Tsingou (FPUT) model discussed below is expected to be in the same universality class of transport as, say, a single-walled nanotube or a nanowire, made of complicated assemblies of atoms. 

Building on this theoretical background, we describe how these features can be 
 exploited to achieve control and enhancement of thermal energy conversion. 
 The simplest case is the one in which  nonlinear interaction can be exploited
to obtain a thermal rectifier. 
Anomalous conduction can be also employed to enhance the conversion of thermal energy into mechanical work, as it occurs in thermoelectricity and thermodiffusion.

\begin{figure}
    \centering
    \includegraphics[width=0.9\textwidth]{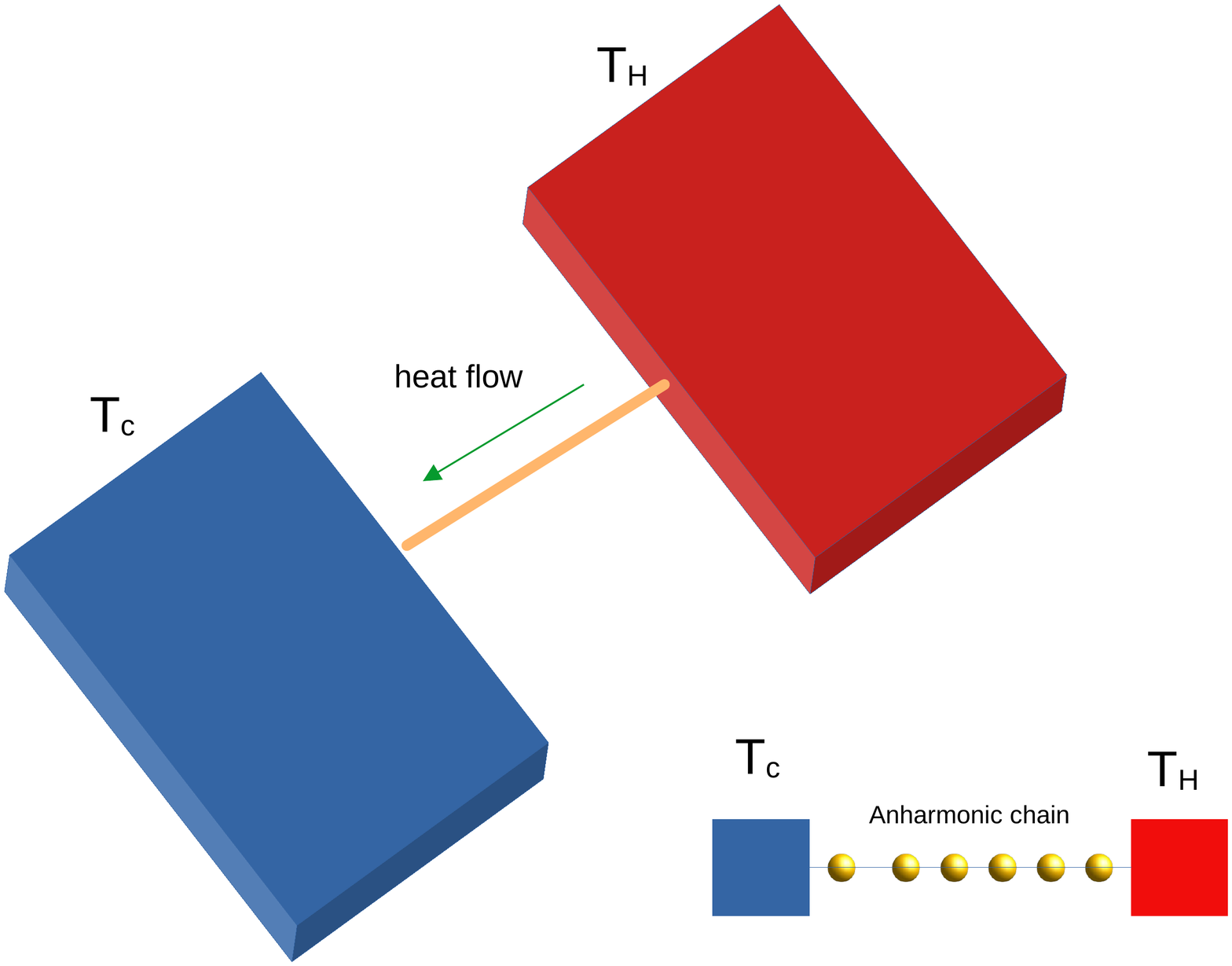}
    \caption{Sketch of a nanoscale heat transfer experimental setup
    to measure the heat conductance of a thin nanosized object in thermal contact with two leads, acting as thermal reservoirs
    at different temperatures $T_H>T_C$.
    Lower figure: a simple idealized model,
    for vibrational energy transport in a 
    one-dimensional structure. It consist
    of an anharmonic chain
    in contacts with two heat reservoirs at different temperatures. The reservoirs can be modeled, for instance, by adding Langevin thermostats at assigned temperatures \cite{LLP03,Dhar08}.
    }
    \label{fig:scheme}
\end{figure}

A crucial issue is the applicability of the theoretical concepts to real nanomaterials. The fabrication of one- and two-dimensional nanomaterials (nanotubes, nanowires, graphene,  two-dimensional semiconductor membranes etc.) is nowadays a concrete experimental possibility. For the sake of illustration, a typical setup we have in mind is sketched in Fig. \ref{fig:scheme}.
Thus one may hope to test experimentally the predictions from statistical physics. In this respect, molecular dynamics (MD) is a fundamental tool to investigate and test theories.
We will provide a review of current research focused on non-Fourier heat transport.

This work complements existing reviews of the research done in the nonequilibrium statistical physics community~\cite{LLP03,DHARREV,dhar2019fourier,benenti2020anomalous} by reporting some recent advances, but also with an eye to applying the concepts to thermal management and to realistic atomic simulations, of interest for condensed-matter.
Although our main focus is on the microscopic foundations,
in this respect, we have to mention that thermodynamic and mesoscopic phenomenological 
approaches have also been worked out to tackle the problem of non-Fourier
heat transport in various materials and devices. 
This leads to various generalization of the heat conduction laws.
For the sake of space, we cannot fully account
for all these contributions in this review: the reader can recover a basic
information about these approaches in the existing literature \cite{sellitto2016mesoscopic,dong2015dynamical,cimmelli2009different,lebon2014heat,kovacs2015generalized,wang2011non}.
For completeness, we mention that reviews in non-Fourier heat transport from a phonon physics viewpoint can be found in \cite{chen2021non} and~\cite{zhang2020size}.
It is now recognized that the reduced dimensionality originates unusual size-dependent features in nanomaterials. Phonon confinement, surface, and interfacial scatterings at the nanoscale are also relevant research topics in this context~\cite{zhang2020size}.  
Before entering the main matter, we give a brief historical account.

\subsection{Historical perspective}

From a historical perspective, one discovers that the problem of non-diffusive heat transport affected quite soon the theoretical foundations of statistical mechanics.  At the end of the XIX century Boltzmann's equation had solved both the equilibrium problem and the hydrodynamics of the ideal gas~\cite{Huang} within a fully consistent approach, which allowed to interpret phenomenological equations, such as the Navier-Stokes and the Heat equation, as direct outcomes of the kinetic theory. The interest for studying the problem of energy (heat) transport in microscopic models of matter, rather than simply relying upon the coarse-grained description of an ideal fluid, emerged quite soon in the XX century. For instance, Peter Debye in the 1930s argued that the problem of heat transport in solids could not be modeled by a harmonic crystal. In fact, the thermodynamics of such a system amounts to a gas of non-interacting quasi-particles, in which the harmonic waves (that now we usually call phonons), propagate ballistically at the speed of sound. 
Accordingly, no diffusion mechanism can be at work in such a model of a solid that  propagates any perturbation ballistically, as a heat superconductor.
The Dutch physicist suggested that {\sl nonlinear interactions} and/or {\sl disorder} should be included in the model, in order to allow for a microscopic mechanism of phonons scattering. His pupil, R. Peierls, approached this problem in the framework of quantum mechanics with phonons being the basic ingredients as quantized linear waves. 
Taking advantage of the arguments raised by Debye, he built up a phenomenological quantum theory of heat transport in solids by introducing the so-called {\sl umklapp processes}, i.e. effective interactions among phonons, treated as a perturbative first-order correction of the unrealistic harmonic limit (more precisely, they correspond to phonon scattering processes that do not conserve the quasi-momentum, see \cite{Peierls}).
This "trick" was able to restore diffusion as the basic mechanism yielding energy transport, as expected also in classical models. 

The seminal considerations raised by Debye certainly inspired also E. Fermi to pursue the other possibility, i.e. including explicitly the role of nonlinearity in classical models. Fermi could tackle this problem because in the last part of his life he and his collaborators (J. Pasta, S. Ulam and M. Tsingou) had access to the biggest computer facility of the time, the MANIAC digital computer in Los Alamos, designed by J. von Neumann. They used MANIAC to perform the first numerical simulation of the dynamics of a 1D anharmonic crystal. Fermi expected that the presence of nonlinear interactions could allow any out-of-equilibrium initial condition to relax spontaneously to thermodynamic equilibrium, signaled by energy equipartition among the normal (Fourier) modes (a more appropriate denomination of phonons in a classical system). 
Contrary to his expectations, they found that the energy of the out-of-equilibrium initial condition (typically a single Fourier mode) initially spreads its energy to other normal modes, but later on, exhibits a quasi-periodic recurrence to the initial state. Nowadays we know that this peculiar phenomenon is due to the presence of soliton-like waves (described by the Korteweg-deVries partial differential equation) in the low-energy limit of the model, which corresponds to the range of parameters explored by Fermi and coworkers. 
Fermi was really puzzled by having not met what he had reasonably conjectured and the account of these numerical simulations appeared only in 1954, the same year when Fermi passed away, just as an internal Los Alamos report~\cite{Fermi1955}.
It is worth pointing out that the quasi-periodic recurrence of the initial states not only challenged the spontaneous evolution to thermodynamic equilibrium, but it was also incompatible with the presence of any diffusive mechanism for energy transport in the FPUT model.  
To make a long story short, we can just mention that energy equipartition was eventually observed in numerical experiments (performed by computers definitely much more powerful than MANIAC), exploring the FPUT model for sufficiently large values of the energy and for a long enough time ~\cite{Chirikov}.

Still following a historical pathway, we have to mention that a complete mathematical description of the peculiar features of a harmonic chain, conjectured by P. Debye, was provided in a seminal paper by Z.Rieder, J.L. Lebowitz and E. Lieb \cite{RLL67}, who obtained an explicit solution of the model of a chain of harmonically coupled oscillators (with fixed boundary conditions)
in contact at its boundaries with stochastic Langevin heat baths at different temperatures $T_+$ and $T_-$. 
A clear signature of ballistic energy transport is testified by an essentially flat temperature profile in the bulk of the chain at $T= (T_+ +T_-)/2$, which decays abruptly at the bath temperatures very close to the corresponding boundaries. Moreover, the heat flux is found to be proportional to the temperature difference, rather than to the temperature gradient. 

This paper provided inspiration to other applied mathematicians for exploring the other suggestion by P. Debye, i.e. including disorder as an ingredient of the model.
The simplest way amounts to introducing isotopic disorder in the harmonic chain by assigning random masses (selected by some probability distribution function) to each harmonic oscillator. 
Relying upon a naive extrapolation of {\sl localization theory}, one could guess that isotopic disorder should turn the {\sl homogeneous superconductor} into a { \sl perfect insulator}, because any propagating energy perturbation should eventually localize, due to the presence of one-dimensional disorder. 
This guess is not entirely correct, because it was found that the mechanism of heat transport in an isotopically disordered chain crucially depends on the adopted boundary conditions 
{One should always take into account that, at variance with equilibrium properties, out-of-equilibrium ones typically exhibit a strong dependence on the boundary conditions.}.
A preliminary important achievement was the rigorous proof that the harmonic chain with isotopic disorder evolves  to a unique stationary non-equilibrium state~\cite{casher1971heat,OConnor74}.
Summarizing a long story (for details see \cite{LLP03}), it was also proved that for {\sl fixed boundary conditions} the heat conductivity $\kappa$ of the disordered harmonic chain of size $N$ vanishes in the thermodynamic limit as
\begin{equation}
\kappa \sim  \frac{1}{\sqrt{N}} \, ,
\end{equation}
while for {\sl free boundary conditions} it diverges in the thermodynamic limit as
\begin{equation}
\kappa \sim  \sqrt{N}\,.
\end{equation}
Accordingly, the latter case shows that anomalous conductivity already emerges in a
disordered harmonic chain with free boundary conditions, thus indicating that despite the
presence of isotopic disorder, superdiffusive heat transport already emerges in such conditions. 

Almost in the same period of time when these contributions about disordered harmonic chains appeared, another community of researchers, concerned with the study of hydrodynamic models, pointed out that anomalous hydrodynamic behavior is a distinctive feature of real fluids \cite{alder1967velocity,dorfman1967difficulties}, through the manifestation of peculiar memory kernels characterizing their space-time correlation functions, and, accordingly, the associated transport coefficients through the Green-Kubo formula (see \cite{livi2017nonequilibrium}). 
Conversely, the attempt of providing a general theoretical approach to the problem of heat transport in fluid dynamics (see \cite{Pomeau1975}) pointed out that superdiffusive (i.e., anomalous) transport properties should characterize one- and two-dimensional fluids - an outcome that, at the time, appeared almost as a mathematical curiosity, in the absence of experimental tests to be performed on real physical systems. 
Nowadays, we are facing the possibility of exploring predictions originated by suitable mathematical models in the realm of nanophysics, where such effects are sizable and of primary interest for material science, as widely reported
in this review article.

The last ingredient invoked for reconciling normal transport properties with low-dimensional systems was deterministic chaos. The basic conjecture was that models of many-body systems exhibiting chaotic dynamics should generate spontaneously diffusive-like behavior of energy perturbations, regardless of their dimensionality, and any other peculiar feature of the model at hand. In fact, the first numerical study of the 1D chaotic model known as Ding-a-ling \footnote{A chain of particles attached to on-site harmonic springs and interacting between themselves by collisions with nearby particles.} confirmed the above
conjecture~\cite{Casati84}. 
Nowadays, we know that normal transport properties in this chaotic model are just a mere consequence of the presence of the on-site force acting on each particle, which prevents the possibility of total momentum conservation, because translation
invariance is broken by the local harmonic forces.  
In fact, contrary to what was conjectured about chaotic systems, when the total length, the total momentum and the total energy of any 1D and 2D chaotic system are conserved anomalous transport sets in. This is what we are going to discuss broadly in Section~\ref{sec:nonequi}.

\subsection{Outline of the paper}

In this work, we will provide an updated review of the research on the problem of anomalous transport, with an eye to its effect on nanoscale systems.
The choice of topics reflects of course our own work and expertise, and we refer the reader to the Bibliography for a more complete overview.

In Section \ref{sec:nonequi} we briefly introduce the main features as studied for simple models, namely low-dimensional lattices of coupled nonlinear oscillators. We recap the relevant phenomena and numerical evidence of anomalous energy transport and emphasize their universal features.
Some recent results, including the effects of magnetic fields, long-range forces, and weak chaos will be reviewed. To demonstrate the generality of these results, we also report on some simulation studies of low-dimensional fluids.

Section \ref{sec:manag} addresses the general problem of how to control and manage heat transfer in  nanosized structures, exploiting the general knowledge gained by recent research. The section deals first with thermal rectification and possible enhancement mechanisms of heat-to-work energy conversion.
Then we review the counterintuitive phenomenon of inverse coupled currents, whereby currents flow against both temperature and concentration gradients.

As a step towards implementation of the above concepts in realistic low-dimensional structures, in Section~\ref{sec:atom} we review the current state of the art of molecular dynamics simulations of nanosystems that are the best candidates to observe and exploit anomalous heat conduction in devices. 
We first recall the basic simulation methods, which can be grouped into equilibrium and non-equilibrium molecular dynamics approaches.
We then describe anomalous transport in two cases: isolated single polymers,
individual carbon nanotubes, and graphene. These appear to  be ideal candidates to test the theoretical predictions in one and two dimensions. 
Then we discuss the relevance of hydrodynamic effects on non-Fourier heat transport at the nanoscale.
Finally, we review some of the works in which the original concept of thermal rectification based on phonons non-linearity has been investigated.

A few experimental works have been reported to try to  validate the theoretical prediction.
In Section \ref{sec:exp} we briefly point to the relevant literature.

\section{Statistical mechanics of non-Fourier heat transport}
\label{sec:nonequi}
In this Section, we review the main features of anomalous  energy transport and violations of Fourier law in low-dimensional non-linear systems from the point of view of statistical physics.
For a more extensive account we refer the reader to existing review papers~\cite{LLP03,DHARREV,Lepri2016,benenti2020anomalous}. 

Anomalous transport emerges as a hydrodynamic effect due to the conspiracy of reduced space-dimensionality, conservation laws, and non-linear interactions, yielding nonstandard relaxation properties even in a linear response regime. The most concise, heuristic way to describe what anomalous energy transport is about, is to imagine that the energy carriers (e.g. phonons in a low-dimensional structure) propagate \textit{superdiffusively}, i.e. in a non-Brownian way.
Indeed, most of the anomalous transport phenomenology in such many-body systems can be effectively described as a random propagation \cite{Zaburdaev2015} of the energy carriers, as demonstrated extensively in the literature~\cite{Cipriani05,Lepri2011,Dhar2013}.
A L\'evy walk is a stochastic process where a particle  undergoes random jumps with constant speed and for random times distributed as an inverse power-law distribution \cite{Zaburdaev2015}.
This description entails that the mean free path becomes effectively infinite, thus leading to the breakdown of standard kinetic approaches. 

\subsection{Anomalous heat transport in classical anharmonic chains} 

As a reference system, we will consider the class of models, represented by a Hamiltonian of the following form:
\begin{equation}
H = \sum_{n=1}^N \left[\frac{p_n^2}{2m} + U(q_{n+1} - q_n) \right]\,.
\label{Hamil}
\end{equation}
The model admits two physical interpretations: one as a discretization of a scalar field $q(x,t)$ on a one-dimensional lattice, $q_n(t)\equiv q(n,t)$.
Otherwise, one can regard $q_n$ as physical positions on the line, i.e 
as a 1D fluid. One implicitly assumes that particle order is preserved. For isolated chains, one typically enforces periodic boundary conditions $q_{N+1}=q_{1}+L$, where $L$ is the total length. 
The Hamiltonian (\ref{Hamil}) has the invariance property under translations $q_n \to q_n + cst$ leading to momentum conservation. Moreover, the "stretch" variable $\sum_n(q_{n+1}-q_n)$ is also conserved. 
Since we are interested in heat transport, one can set the total momentum to zero without loss of generality. The relevant state variables of microcanonical equilibrium are the specific energy (i.e., the energy per particle) $h=H/N$ and the  elongation $l=L/N$ (i.e., the inverse of the particle density).
If $U$ is quadratic around the equilibrium state,  the chain admits a single branch of acoustic phonons. 

Common choices for the interaction potential are the famous FPUT potential 
\begin{equation}
U_{FPUT}(x) \equiv \frac{1}{2} x^2+\frac{\alpha}{3} x^3 + \frac{\beta}{4} x^4  
\label{eq:fput}
\end{equation}
or the Lennard-Jones one \cite{Lepri2005}
\begin{equation}
U_{LJ}(x) \;=\;  \frac{1}{12}
\bigg(\frac{1}{x^{12}}\, -\, \frac{2}{x^6}\,+\, 1 \bigg).
\label{lj}
\end{equation}
Throughout this paper, we adopt suitable adimensional model units to 
simplify the notation.

The main results, emerging from a long series of contributions, can be summarized as follows \cite{LLP03,DHARREV,Lepri2016,benenti2020anomalous}. Models of the form (\ref{Hamil}) 
having the three conservation laws described above,
generically display  
\textit{anomalous} transport and relaxation features.
In other terms, 
Fourier's law \textit{does not hold}: the kinetics of energy carriers is so
correlated that they are able to propagate \textit{faster} than in the ordinary (diffusive) case. 

Such a behaviour manifests itself in the simulations in 
various facets \cite{lepri2016heat}:
\begin{itemize}
\item The finite-size heat conductivity $\kappa(L)$  diverges  in
the limit of a large system size $L\to \infty$ as , 
$\kappa(L) \;\propto\; L^\gamma$  \cite{LLP97}, 
i.e. the heat transport coefficient is ill-defined in the thermodynamic limit.

\item The equilibrium correlation function of the total
energy current $J$ displays a nonintegrable long-time tail 
$\langle J(t)J(0)\rangle \propto t^{-(1-\delta)}$, with $0\le\delta < 1$  
\cite{Lepri98a,L00}. Accordingly, 
the Green-Kubo formula yields an infinite value of the conductivity.\footnote{An alternative formulation of this statement, based on the  well-known Wiener-Khintchine theorem and asymptotic analysis, is that the equilibrium power spectrum of the energy current has a zero-frequency singularity of the form $\omega^{-\gamma}$.}

\item Energy perturbations propagate superdiffusively \cite{Denisov03,Cipriani05}: a local disturbance of the energy field spreads, while  its variance broadens in time as
$\sigma^2\propto t^{\beta}$,
with $\beta > 1$.

\item Temperature profiles in the nonequilibrium steady states 
are intrinsically nonlinear, even for vanishing applied temperature 
gradients. Typically they are the solution of a \textit{fractional heat equation} instead of
the standard heat equation
\cite{Lepri2011,Kundu2019,Dhar2019}.

\end{itemize}

There is a large body of numerical evidence that the above features
occur generically in 1D, whenever the conservation of energy, 
momentum and length hold. As it is known, 
this is related to the existence of
long-wavelength (Goldstone) modes  (an acoustic phonon branch in the linear
spectrum) that are very weakly damped.
Moreover, the exponents that characterize the phenomena
listed above are related, as they are different facets of the 
same physical effects. For instance, if there is a finite propagation
speed one can argue that $\delta=\gamma$, etc. \cite{LLP03,DHARREV,Lepri2016,benenti2020anomalous}.

As for the dependence on the dimension, one can find evidence of anomalous 
transport and a diverging heat conductivity 
in 2D model, like for instance a 2D version of the 
the FPUT model. Typically, the finite-size conductivity 
data are consistent with a 
logarithmic dependence with the system size $L$ \cite{Lippi00,Wang2012,DiCintio2017}.
Moreover, normal diffusive transport is restored 
in the 3D case \cite{Saito2010a,wang20103D} (with the exception of
integrable models like a purely harmonic crystal). 

The importance of conservation laws can be appreciated by
examining the following examples. 
Consider first the  anharmonic chains where a local 
external pinning potential $V$ is added to the Hamiltonian,
\begin{equation}
H = \sum_{n=1}^N \left[\frac{p_n^2}{2m} + U(q_{n+1} - q_n) + V(q_n) \right]\,.
\label{pinned}
\end{equation}
The main difference is that this class of models has only one global
conserved quantity, namely energy. The dispersion in the linear 
limit is gapped, i.e. it admits only optical phonon modes. 
Relevant examples in this class that have 
been discussed in the present context are the discrete 
$\phi^4$ theory \cite{Aoki00} and the Frenkel-Kontorova model \cite{Hu1998}. 
Generally, the addition of pinning forces suffices to make the anomalies disappear and restore Fourier's law and standard heat diffusion.
Another interesting and well-studied example~\cite{Giardina99,Gendelman2000} is the  rotor (or Hamiltonian XY) chain, namely (\ref{Hamil}) with 
$$
U(x)= U_{XY}(x) \equiv 1-\cos x.
$$
Here, $x$ has to be read as a difference between angles, and thus the 
stretch variable is no longer globally conserved by a change 
of relative phases by $2\pi$. 
So the model has two conserved quantities only, and transport is normal with finite thermal conductivity and finite Onsager coefficients  \cite{Iacobucci2011,Iubini2016,spohn2014rotors}.
(A more sound theoretical justification for those observations
will be reviewed below).

For completeness, we also mention 
another important model related to the ones
mentioned above, namely the 
discrete nonlinear Schr\"odinger (DNLS) lattice 
(also termed the discrete Gross-Pitaevskii equation). 
As it is known, it has important applications in many domains, for instance, electronic transport in biomolecules  and 
atomic condensates in optical lattices \cite{Kevrekidis}. It is defined by the Hamiltonian
\begin{equation}
H=\sum_{i=1}^{N}\left[\frac{1}{4}\left(p_i^2+q_i^2\right)^2+
p_ip_{i+1}+q_iq_{i+1} \right],
\label{eq:dnls}
\end{equation}
whose equation of motion can be written in terms of the complex variable $z_n = (p_n + i q_n)/\sqrt{2}$ as
\begin{equation}
\label{eq:z}
i \dot{z}_n=-2\mid z_n\mid ^2 z_n -z_{n-1}- z_{n+1}.
\end{equation}
Besides the total energy, the system admits a second constant of motion, namely the total norm $A = \sum_{n=1}^N \mid z_n \mid^2 $, which, depending on the physical context, can be interpreted as the gas particle number, optical power, etc.
At variance with its continuum counterpart, the DNLS is non-integrable and typically displays chaotic dynamics. There is evidence that transport is normal for DNLS, i.e. that the elements of the Onsager matrix are well-defined in the 
thermodynamic limit
\cite{Iubini2012,Iubini2013a,Mendl2015,iubini2017chain}. 

To end this section, let us mention that the actual simulations described here turned out to be quite challenging, despite the apparent simplicity of the models. As it is known, equilibrium simulations of correlation functions are affected by large statistical fluctuations. Reliable measurements of transport coefficients in nonequilibrium simulations can also be very difficult, while the  presence of  boundary resistances and strong finite-size effects are usually serious issues. It is worth mentioning that improved methods, which originated in the statistical physics community, seem very promising. For instance, 
importance sampling schemes aiming at sampling the probability of rare current fluctuation have been proposed~\cite{gao2019nonlinear}. This approach, based on the so-called cloning algorithms, seems quite effective in assessing the anomalous transport properties of low-dimensional nanomaterials \cite{ray2019heat}.

\subsection{Kardar-Parisi-Zhang universality}

The main theoretical insight that has been achieved is the intimate relation between the anharmonic chain and one of the most important equations in nonequilibrium statistical physics, the celebrated Kardar-Parisi-Zhang (KPZ) equation, also known as the noisy (or fluctuating) Burgers equation.
The latter was originally introduced in the (seemingly unrelated) context of surface growth \cite{barabasi1995fractal} but turned out to be a paradigm of many non-equilibrium problems in physics. In the case of a scalar stochastic field $h(x,t)$ in one spatial dimension it reads
\begin{equation}
\frac{\partial h}{\partial t}=
\nu\frac{\partial^2 h}{\partial x^2}+
\frac{\kappa} 2\left(\frac{\partial h}{\partial x}\right)^2 
 +  \eta,
\label{kpzh}
\end{equation}
where $\eta(x,t)$ represents a Gaussian white noise with $\langle\eta(x,t)\eta(x',t')\rangle$=$2D\delta(x-x')\delta(t-t')$ and $\nu,\kappa,D$ are the relevant parameters. 

The connection between the KPZ and anomalous transport has been derived within the nonlinear fluctuating hydrodynamics approach \cite{Spohn2014,VanBeijeren2012}. 
The theory is able to justify and predict several universal features of anomalous transport in anharmonic chains. This implies that their large-scale dynamical properties are in the same dynamical universality class as Eq.(\ref{kpzh}).
We give here a preliminary account, referring the reader to 
\cite{Spohn2014,spohn2016lnp} for details. 
The main entities are  the random fields describing deviations of the conserved quantities  with respect to their stationary values. The role of fluctuations is taken into account by the renormalization group or some kind of self-consistent theory.

More precisely, large-scale fluctuations result from three stochastic fields or modes: two sound modes, $\phi_\pm$, traveling at speed $c$ in opposite directions, and one stationary but decaying heat mode, $\phi_0$.
Loosely speaking, we can represent e.g. the displacement field as the superposition of counter-propagating plane waves, modulated by an envelope that is ruled at large scales by Eq.~(\ref{kpzh}).
The quantities of interest are the equilibrium spatiotemporal correlation functions $C_{s s'}(x,t)=\langle \phi_s(x,t) \phi_{s'}(0,0)\rangle$, where $s,s'=-,0,+$. Because the modes separate linearly in time, one argues that the corresponding equations decouple into three single-component equations. 
Those for the sound modes have precisely the structure of the noisy Burgers equation \cite{Spohn2014}.
For the heat mode, the self-coupling coefficient vanishes, whatever the interaction potential and sub-leading corrections, must
be considered within the mode-coupling approximation, resulting in the symmetric L\'{e}vy-walk distribution. 

For the generic case of non-zero pressure, which corresponds either to asymmetric inter-particle potentials or to an externally applied stress, the theory predicts the following scaling form for the auto-correlation functions of the modes  
\begin{align} 
  C_{\mp\mp}(x,t)&=   \frac{1}{(\lambda_s t)^{2/3}}~ f_{\mathrm{KPZ}} 
  \left[~ \frac{x \pm ct}{(\lambda_s t)^{2/3}}~\right]~, \label{eqscalS}\\
  C_{00}(x,t) &=  \frac{1}{(\lambda_e t)^{3/5}} ~f^{5/3}_{\mathrm{LW}}\left[~ \frac{x}{(\lambda_e t)^{3/5}}~\right]~.\label{eqscalE} 
\end{align}
Remarkably, the scaling function $f_{\rm KPZ}$ is universal and known exactly (see \cite{Spohn2014} and references therein). 
Also, $f_{\mathrm{LW}}^\nu(x)$ denotes the L\'{e}vy function
of index $\nu$,  
defined as the Fourier transform of the characteristic function $e^{-\mid k \mid^\nu}$. Such a family of functions is well-known
and (not by chance) appears in the theory of 
anomalous diffusion \cite{AnotransBook08}.
The non-universal
features are in the $\lambda_s$ and $\lambda_e$, that are  model-dependent parameters.
Therefore, fluctuations of observables display, in the hydrodynamic
limit, \textit{anomalous dynamical scaling}, as 
manifested by the $t^{1/z}$ dependence of correlation
on time, as seen in Eq.\ref{eqscalS}. Here
the \textit{dynamical exponent} $z=3/2$ is different from $z=2$, expected for a standard diffusive process. 
\footnote{Equivalently, one can 
recast the results in reciprocal space. For instance, the dynamical 
structure factor $S(k,\omega)$ of the particle displacement shows 
for $k\to 0$ two sharp peaks at $\omega = \pm ck$
that correspond to the propagation of sound modes 
and for $\omega\approx\pm \omega_{\rm max}$ behave as
\begin{equation}\label{scaling32}
S({k},\omega)\sim \hat{f}_{\rm KPZ}\left(\frac{\omega\pm\omega_{\rm max}}{\lambda_s k^{z}}\right),
\end{equation}
with $\hat{f}_{\rm KPZ}$ being the transform of the scaling function. 
}
As a consequence of such scaling, it has been argued that 
the energy current correlations decay with an exponent 
$\delta=1/3$ i.e. $\kappa(L) \sim L^{1/3}$.

Even from such a short outline, it is evident that the hydrodynamic approach provides several predictions.
Most of them have been successfully tested for several models~\cite{Mendl2013,Das2014,Mendl2014}, including anharmonic chains with three conserved quantities like the FPUT model \cite{Das2014a,DiCintio2015}. 
Also, the KPZ scaling is not limited to scalar displacements but applies as well to quasi-1D chains with 3D motions \cite{Barreto2019}.
It should be however recognized that departures from universality  have been reported in simulations of specific hard-point gas models \cite{Hurtado2016}. 

Another remarkable feature is that the KPZ scaling may hold approximately in some regimes. For instance, it accounts well for the low-temperature hydrodynamics of the DNLS equation (\ref{eq:z}) in the regime where there is an approximate third conservation law \cite{Kulkarni2015,Mendl2015}. It should however be kept in mind that the observability of KPZ scaling may be hindered by significant scale effects, especially when the dynamics is only weakly chaotic \cite{lepri2020too} (we will come back to this in Subsection \ref{sec:integ} below).

Although the KPZ class is expected to be generic, there may be specific cases that belong to a different (non-KPZ) universality class by virtue of additional symmetries. This is the case of anharmonic chains models with symmetric interaction potentials like the FPUT-$\beta$ model, namely the potential in Eq.~(\ref{eq:fput}) with $\alpha=0$ \cite{Lepri03,Lee-Dadswell2015}, as well as of chains with conservative noise (e.g., the noisy harmonic \cite{BBO06,Lepri2009} or nonlinear \cite{Basile08,Iacobucci2010} chains). 
Physically, this condition corresponds to the case where the chain is at zero pressure \cite{Lee-Dadswell2015}.
This possibility is also accounted for by the fluctuating hydrodynamics approach. Indeed, for an even potential, the mode-coupling approximation predicts a different scaling of the correlation functions \cite{Spohn2014}:
\begin{align} 
 C_{\mp\mp}(x,t)&= \frac{1}{{(\lambda^0_s t)}^{1/2}} f_{\mathrm{G}} \left[~ 
 \frac{x \pm ct}{(\lambda_s^0 t)^{1/2}}~\right]~,\label{eqscalSP
0}\\
  C_{00}(x,t) &=  \frac{1}{(\lambda_e^0 t)^{2/3}} ~f^{3/2}_{\mathrm{LW}}\left[~ \frac{x}{(\lambda_e^0 t)^{2/3}}~\right]~,\label{eqscalEP0}
\end{align} 
where $f_\mathrm{G}(x)$ is the unit Gaussian with zero mean. This class would correspond to a diverging finite-size conductivity with $\gamma=1/2$.
It is worth also mentioning, that the kinetic theory approach confirms a non-KPZ behavior for the FPUT-$\beta$ model \cite{Pereverzev2003,Nickel07,Lukkarinen2008,dematteis2020coexistence}, albeit with a different exponent $\gamma=2/5$, which is pretty close to the numerical one \cite{Lepri03,Wang2011}.

To conclude, it has been also argued that the two main nonequilibrium universality classes, the diffusive and KPZ, are only two cases of an infinite discrete family. The members of this family can be identified by their dynamical exponent, which depends on both the number of conserved quantities and the coupling among their hydrodynamic modes \cite{Popkov2015}.

\subsection{Anomalous transport with broken time-reversal}
\label{sec:magn}

The effect of time-symmetry breaking on anomalous transport has been considered in~\cite{Tamaki2017,Saito2018} that investigated a one-dimensional chain of charged beads, interacting with  a quadratic potential, in the presence of an external magnetic field ${\bf B}$. 
For ${\bf B}=0$, $\delta = \frac{1}{2}$, thus indicating the presence of anomalous transport and a divergent heat conductivity $\kappa (L) \sim L^{\frac{1}{2}}$.
For nonvanishing ${\bf B}$ the total pseudo-momentum is conserved, and the hydrodynamics of the model must be modified. Actually, numerical and analytic estimates indicate that in this case, the exponent $\delta$ may turn to a value different from  $\frac{1}{2}$.
In particular, in~\cite{Tamaki2017} two different cases were considered: the one where oscillators have the same charge and the one where oscillators have alternate charges of sign $(-1)^n$, $n$ being  the integer index numbering oscillators along the chain. 
In the former case, the sound velocity vanishes, and the energy correlator exhibits a heat peak centered  at the origin and spreading in time. Conversely, in the latter case, the sound velocity is finite and its value depends on ${\bf B}$.
Here, the heat mode is coupled to sound modes propagating through the chain. 
The exponent $\gamma$ is the same obtained for ${\bf B} = 0$, i.e. $\gamma = \frac{1}{2}$.

In the case of equally charged beads, one finds a novel exponent $\gamma =\frac{3}{8}$, which corresponds to a different universality class.
An important remark on this new exponent is that, in the absence of 
a finite sound velocity, the identification of the exponents $\delta$ and $\gamma$ invoked above, no longer holds. Indeed, in this case one has $\delta\approx\frac{3}{4}$.
Rigorous estimates  of these exponents also for the $d=2$ and $d=3$ versions of the model have been obtained through the asymptotes of the corresponding Green-Kubo integrals, where the deterministic dynamics has been  substituted with a stochastic version that conserves the same quantities \cite{Saito2018} . 

\subsection{Integrable models and their perturbations}
\label{sec:integ}

The above results are mostly obtained in a strongly nonlinear regime or more generally far from any integrable limit. For the FPUT model, this means working with high enough energies/temperatures to avoid all the difficulties induced by quasi-integrability and the associated slow relaxation to equilibrium.

Integrable systems constitute \textit{per se} a relevant case and experienced a renovated interest in recent years.  In the framework of the present work, the reference example is certainly the celebrated 
Toda chain, namely the model in Eq.~(\ref{Hamil}) with  
\[
U_T(x) = e^{-x} + x -1\,.
\] 
As intuitively expected, heat transport is ballistic due to its integrability and the associated solitonic solution \cite{Toda79}. 
Mathematically, this is expressed by saying that there is a non-vanishing \textit{Drude weight}, namely, a zero-frequency component of  energy, current, and power spectra \cite{Zotos02,Shastry2010}.
A lower bound of the Drude weight can be estimated by making use of Mazur inequality \cite{Mazur1969} in terms of correlations between the currents themselves and the conserved quantities.
See~\cite{Zotos02,Shastry2010} for application to the Toda case.
However, the idea of solitons transporting energy as independent particles is somehow too simplistic.
At variance with the harmonic chain, which is also integrable but whose proper modes are non-interacting phonons,
the Toda chain should rather be considered as an interacting integrable system \cite{Spohn2018}.
In simple terms, it means that the quasi-particles (the famous Toda solitons) experience a stochastic sequence of spatial shifts as they move through the lattice, interacting with other excitations without momentum exchange. 
This yields a kind of non-dissipative diffusion \cite{Theod1999} that reflects in the calculation of the transport coefficients by the Green-Kubo formula, indicating the
presence of a finite Onsager coefficient. The latter corresponds to a diffusive process on top of the dominant ballistic one \cite{Shastry2010,Kundu2016,DiCintio2018}.

In general, equilibrium correlations of integrable models should display a ballistic scaling, as indeed confirmed for the Toda chain \cite{Kundu2016}. There are some exceptions to this rule, like in the case of the scaling behavior of the integrable lattice Landau-Lifshitz spin chain that, for the case with zero mean magnetization, has a scaling function identical to the one obtained for the KPZ equation \cite{das2019kardar}.

A natural question concerns the behavior when a generic perturbation is applied to an otherwise integrable system. 
The first observation is that the actual form of perturbation is relevant. 
For instance, adding a quadratic pinning potential $V(x)=x^2/2 $ to the Toda chain does restore standard diffusive transport, but numerical simulations show that long-range correlations are preserved over relatively long scales \cite{DiCintio2018,Dhar2019}. Moreover, weak perturbations that conserve momentum (and are thus expected to display anomalous transport in the KPZ class) display instead significant deviations \cite{Iacobucci2010} and even diffusive transport over the accessible simulation ranges \cite{Chen2014}. 

The relevant issue regards the typical length scales over which the anomalous transport is restored by the effect of a perturbation. The length-independent flux exhibited by integrable systems is the result of the free displacement of quasi-particles (the integrals of motion, such as solitons) from the hot towards the cold reservoir.
In the vicinity of the integrable limit, as a result of mutual interactions, the quasi-particles acquire a finite and large mean free path $\ell$. A purely ballistic behavior is observed for $L<\ell$. On the other hand, $L>\ell$ is not a sufficient condition to observe a crossover toward the anomalous behavior predicted by the above-mentioned theoretical arguments.
In fact, it is necessary for $L$ to be so long that the {\it normal} flux induced by inter-particle scattering becomes negligible.
Altogether, upon increasing $L$ at fixed $\ell$, one should see a first ballistic regime followed by a kinetic (diffusive) one, until 
eventually, the asymptotic hydrodynamic (anomalous) regime is attained.
The three different regimes are observable only provided the relevant length scales are widely separated.

Based on these heuristic considerations, one may look for a decomposition of heat flux $J(L,\varepsilon)$ as \cite{lepri2020too}
\begin{equation}
J(L,\varepsilon) = J_A(L,\varepsilon) + J_N(L,\varepsilon),
\label{Jtot}
\end{equation}
where $\varepsilon$ measures the perturbation strength
i.e. the distance from the integrable limit, $J_A$ is the anomalous hydrodynamic part, and
$J_N$ is the kinetic contribution, accounting for the energy transported by the weakly interacting quasi-particles. As explained above, for $L\to\infty$,
we expect $J_A \approx L^{\gamma-1}$ with $\gamma=1/3$ in systems belonging to the KPZ class.

Following a kinetic argument~\cite{pitaevskii2012physical}, we argue that $J_N$ must be 
only a function of  $\xi=L/\ell$, which is the ratio expressed in units 
of the mean free path $\ell$, the only relevant scale. Moreover, $J_N$ should display a crossover from ballistic to diffusive regimes depending on $\xi$, namely it should
approach a 
constant for small $\xi$ and  be proportional to $1/\xi$ for large $\xi$. A simple interpolating 
formula would thus be 
\begin{equation}
J_N(\xi) = \frac{j_0}{r+ \xi}\; ,
\label{JN}
\end{equation}
where $r$ is a constant accounting for the boundary resistance \cite{Aoki01} 
and $j_0$ is an additional constant.

Approaching the integrable limit  
the mean free path must diverge, and it is natural to assume 
that $\ell \approx \varepsilon^{-\theta}$, 
where $\theta >0$ is a system-dependent exponent.
As long as $J_A(L,\varepsilon)$ does not display any singularity for $\varepsilon \to 0$ (we return to this point below), we can neglect its dependence on
$\varepsilon$. 
Altogether, Eq.~(\ref{Jtot}) can be approximated
for large $L$ as
\begin{equation}
J(L,\varepsilon) \;\approx \; \frac{c_A}{L^{1-\gamma}} \,+\, 
 \frac{c_N}{L\varepsilon^{\theta}}  \; ,
\label{Jtot2}
\end{equation}
where $c_A$ and $c_N$ are two suitable parameters. Accordingly,
the anomalous contribution dominates only above the crossover length
$\ell_c \approx \varepsilon^{-\theta/\gamma}$. For $L\le \ell_c$, heat conduction is
dominated by $J_N$. In particular, within the range $[\ell=\varepsilon^{-\theta},\ell_c]$
an \textit{apparent} normal conductivity is expected, 
which is nothing but a finite size effect.

The above description accounts very well for the 
numerical data for the perturbed Toda and hard-point
gas models \cite{lepri2020too}. 

The standard case of the perturbed harmonic chain deserves a special consideration
from this point of view. 
Numerical analysis of the FPUT$-\beta$ model at very low energy,
i.e. below the strong stochasticity threshold, does not reveal any signature of
an intermediate diffusive regime, but rather a direct crossover 
from ballistic to anomalous regimes \cite{Lepri05}.
More compelling evidence of the absence of a diffusive regime comes from the study 
of the harmonic chain with conservative noise~\cite{BBO06} with rate 
$\varepsilon \to 0$. 
It has been found analytically~\cite{Lepri2009} and confirmed numerically~\cite{Delfini10} 
that $J_A(L,\varepsilon)$, exhibits a singular dependence 
in the form of a divergence of the coefficient $c_A$ in Eq.~(\ref{Jtot2}), $c_A \approx \varepsilon^{-1/2}$,
which implies that $J_A$ dominates $J_N$ for any value of $L$.

The above reasoning can  explain the numerical
observation of the apparent normal diffusion observed
for asymmetric potentials $U(x)\neq U(-x)$ ~\cite{Iacobucci2010,Chen2014,Zhong2012,Wang2013}. 
Indeed, if the potential is well approximated by a perturbed 
Toda one, the crossover to the anomalous regime may occur at 
prohibitively large sizes. For instance, the paradigmatic FPUT chain 
is consistent with this scenario \cite{Das2014}.
Other studies confirmed that the diffusive regime is indeed a finite-size effect, 
whereby anomalous behavior is recovered
for $L$ large enough \cite{Wang2013,Das2014,Lee-Dadswell2015a,Miron2019}.

\subsection{Long-range interacting chains}

Another ingredient that has been considered in this context
is the effect of long-range interactions, i.e. systems in which the interparticle potential decays at large 
distances $r$ as $ r^{-d-\sigma}$, in dimension $d$. \cite{RuffoRev,Campa2014}.
The study of this class of problems has a long-standing tradition in equilibrium statistical mechanics, starting from the 
seminal works by F.J. Dyson \cite{dyson1969existence}. Besides the 
theoretical motivations, there are also experimental 
example, notably trapped ion chains, dipolar condensates
etc.  both classical and quantum \cite{defenu2021}.

One distinguishing 
feature is that, for interactions decaying sufficiently slowly 
with distance, perturbations may propagate with infinite velocities, yielding qualitative differences with respect to their short-ranged counterparts \cite{Torcini1997,Metivier2014}.
At nonequilibrium, the dynamics of long-range systems presents metastable states, whose lifetime scales as $N$ \cite{Bouchet2010,RuffoRev,Campa2014} and even lack of thermalization upon interaction with a single external bath \cite{deBuyl2013}.
As far as transport and hydrodynamics are concerned, non-local effective equations are expected to arise naturally by the non-local nature of couplings. This has also effects on energy transport for open systems interacting with external reservoirs and, more 
generally, on the way in which the long-range terms couple the system with the environment. 

\begin{figure}
\centering
\includegraphics[width=0.7\textwidth]{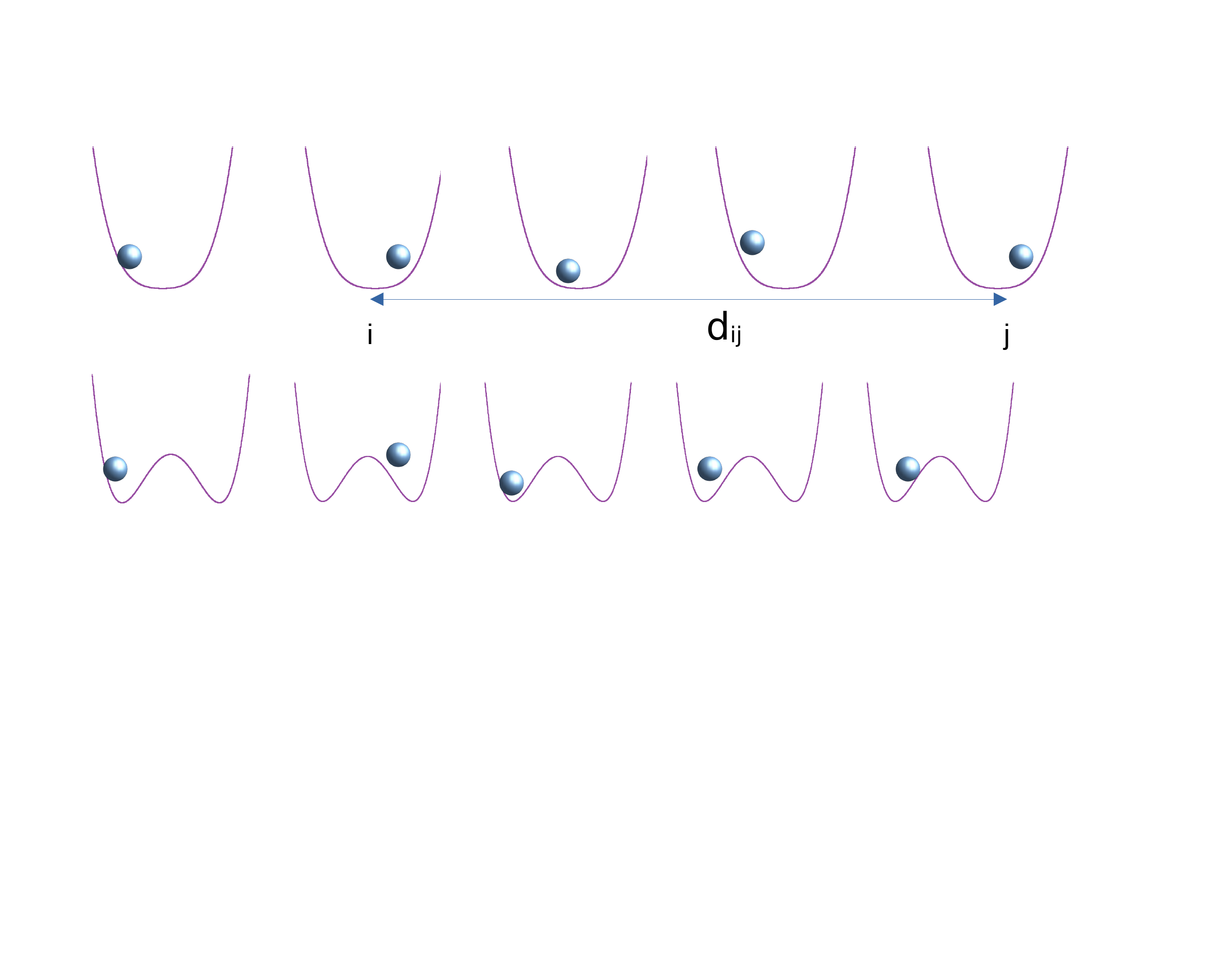}
\caption{Schemes of a one-dimensional coupled-oscillator model
with single and double-well local potential and
long-range interaction decaying as $d_{ij}^{-1-\sigma} $ , $d_{ij}$ represents 
the distance between the lattice sites, and the Hamiltonian 
is given by (\ref{eq:lrphi4}).} 
\label{fig:long}
\end{figure}

For the class of nonlinear oscillator lattices treated here, there is now a body of evidence that 
non-Fourier transport would occur, although several
issues are still open.
Let us first considered the simplest extension of (\ref{pinned})
that contains long-range harmonic couplings (see Fig. \ref{fig:long})
\cite{iubini2021hydrodynamics}
\begin{equation}
H = \sum_{i=1}^N \left[\frac{p_i^2}{2} + V(q_i)-
\frac{\mu}{\mathcal N_\sigma}\sum_{j> i}^{N} \,\frac{q_i q_j}{d_{ij}^{1+\sigma}} \right]
\label{eq:lrphi4}
\end{equation}
where $\mu$ is a coupling constant. 
The cases $\mu>0$ and $\mu<0$ correspond to ferromagnetic (attractive) and antiferromagnetic (repulsive) interactions, respectively. 
For a finite, periodic lattice, $d_{ij}$ identifies the shortest distance between sites $i$ and $j$ 
\begin{equation}
\label{eq:d}
d_{ij}=\min\{\mid i-j\mid,N-\mid i-j\mid\}.
\end{equation}
The real exponent $\sigma\ge-1$ controls the interaction range. In one-dimension, for $\sigma<0$ the energy
is made extensive by the Kac prescription 
~\cite{RuffoRev} i.e. by introducing the factor 
\begin{equation}
\label{eq:kac}
{\mathcal N}_\sigma  =   2\sum_{r=1}^{N/2}\frac{1}{r^{1+\sigma}}.
\end{equation}
For $\sigma > 0$, ${\mathcal N}_\sigma $ attains a constant value for large
sizes $N$ and diverges for $\sigma < 0$, ensuring energy extensivity.
Notice that $\sigma = -1$ corresponds to a mean-field interaction, $N_{-1}=N$ \cite{desai1978statistical,dauxois2003clustering}, 
while in the limit of $\sigma\to+\infty$ the case of nearest-neighbor interactions is retrieved. 

Another class of models is the one with \textit{nonlinear} long-range forces. Two specific examples have been considered
in the literature. The first one is a harmonic 
chain with long-range quartic coupling \cite{Bagchi2017,bagchi2021heat}
\begin{equation}
\label{eq:bagchi}
H = \sum_{i=1}^N \left[\frac{p_i^2}{2} + 
\frac{1}{2}(q_{i+1}-q_i)^2+
\frac{1}{{\mathcal N}_\sigma}  \sum_{j\neq i}^{N} \,
\frac{(q_i-q_j)^4}{d_{ij}^{1+\sigma}} \right].
\end{equation}
The second has instead the form \cite{Olivares2016,Iubini2018}
\begin{equation}
\label{eq:olivares}
H = \sum_{i=1}^N \left[\frac{p_i^2}{2} + 
\frac{1}{{\mathcal N}_\sigma}  \sum_{j\neq i}^{N} \,
\frac{U(q_i-q_j)}{d_{ij}^{1+\sigma}} \right],
\end{equation}
with $U=U_{XY}$ \cite{Olivares2016} and $U=U_{FPUT}$ \cite{Iubini2018}.
The two models differ in their
dispersion relation in the harmonic limit, as obtained by seeking for plane-wave solutions of the form $q_n\sim\exp(\imath kn-\imath \Omega_\sigma t)$
(for periodic boundary conditions the allowed values of the wave number $k$ are integer multiples of $2\pi/N$). 
In fact, model (\ref{eq:bagchi}) has the standard, $\sigma$-independent acoustic dispersion $\Omega^2(k)=4\sin^2(k/2)$ and finite group velocities.
Instead, for (\ref{eq:olivares}) the dispersion relations read 
\cite{Miloshevich2015,Chendjou2018} :
\begin{equation}
\Omega_\sigma^2(k) = \frac{2}{\mathcal{N}_\sigma} \sum_{n=1}^N \frac{1-\cos kn}{n^{1+\sigma}}\,.
\label{lindisp}
\end{equation}
For $N\to \infty$ and in the small
wavenumber limit, $\mid k \mid\to 0$,
\begin{equation}
\Omega_\sigma(k) \propto
\begin{cases}
\mid k\mid^{\frac{\sigma}{2}} & \text{for} \quad 0<\sigma<2,\\
\mid k\mid &\text{for} \quad \sigma\ge 2.
\end{cases}
\label{lindisp2}
\end{equation}
As a consequence, the group velocity diverges as $\mid k\mid^{\frac{\sigma-2}{2}}$
in the first case, while it is finite in the second one. This result can 
also be 
derived from the continuum limit, yielding a fractional wave equation \cite{Tarasov2006}.
The case $\sigma<0$ is even more peculiar, since 
the spectrum remains discrete in the 
thermodynamic limit \cite{defenu2021metastability} and will not be considered henceforth.

Let us first discuss the case (\ref{eq:lrphi4}), and focus on the so-called weak-long range 
case, $\sigma>0$. As far as hydrodynamics is 
concerned, the main results can be described effectively 
assuming that the fluctuations of the local energy field
\begin{equation}
h_i = \frac{p_i^2}{2} + V(q_i)-
\frac{\mu}{{\mathcal N}_\sigma }\sum_{j\neq i}^{N} \,\frac{q_i q_j}{d_{ij}^{1+\sigma}}, 
\end{equation}
propagates as a L\'evy flight. This process is well-known as the simplest generalization
of the Brownian random walk, yielding anomalous diffusion of an individual particle \cite{metzler2000random}.
More precisely, we assume that the site energies $h_j$ undergo a 
stochastic process ruled by the master equation 
\begin{equation}
\dot h_j = \sum_{i\neq j} W_{ij}(h_i-h_j), \qquad
W_{ij}=\frac{\lambda}{\mid i-j \mid^{\sigma +2}},
\label{eq:ME}
\end{equation}
where $\lambda$ is a characteristic rate, setting the inverse timescale of the 
process. Remarkably, this simple model accounts both for the dynamical scaling of structure factors and for the scaling of the energy flux out-of-equilibrium \cite{iubini2021hydrodynamics}.
Large-scale fluctuations of the local energy field display hydrodynamic 
behavior, which is diffusive for $\sigma>1$ and superdiffusive for $0<\sigma<1$
in both the cases with single and double-well
local potentials, with either attractive or repulsive couplings.
In the superdiffusive case, numerical data 
and (\ref{eq:ME}) suggest that the 
energy field follows  
a fractional diffusion equation of 
order $\sigma$, i.e. a non-Fourier heat transport 
with an anomalous scaling of the energy flux as
$L^{-\sigma}$. Remarkably,  
in the case of the double-well potential with attractive interaction
such behavior of energy fluctuations appears to be insensitive to the 
phase transition.

In hindsight, the very fact that a L\'evy flight model account
for the large-scale energy fluctuations and transport may
appear an obvious consequence that the couplings 
decay algebraically with the distance. However 
this is not the case for at least two reasons. 
First of all, the correct exponent in (\ref{eq:ME}) must be $\sigma+2$, 
which is not trivial a priori. Second, models with the same   $r^{-1-\sigma}$ interactions
like (\ref{eq:bagchi}) and (\ref{eq:olivares})  have 
different dynamical exponents \cite{DiCintio2019,Olivares2016,Bagchi2017}.
For instance for the FPUT model,  
the structure factors at finite energy density display distinct peaks,
corresponding to long-wavelength propagating modes, whose dispersion relation is compatible
with (\ref{lindisp2}). Also, they display 
dynamical scaling of the form (\ref{scaling32}) , 
with a dynamical exponent $z$ that depends weakly on $\sigma$ in the range 
$0<\sigma<2$. The lineshapes have a non-trivial 
functional form and appear somehow independent of $\sigma$. 

In other words, models having the same coupling $r^{-1-\sigma}$
may belong to different dynamical universality classes, having
different hydrodynamics. 
This conclusion is confirmed also by noting the differences between model (\ref{eq:lrphi4})
and the momentum-exchange model
with long-range interactions \cite{tamaki2020energy}.
It consists of 
the Hamiltonian (\ref{eq:lrphi4}) with  
a quadratic pinning potential $V$ perturbed by a random exchange of momenta between the
nearest neighbor sites, occurring with a given rate \cite{tamaki2020energy}.
This model allows for an exact calculation of the exponent $\delta$ ruling the decay of energy current autocorrelation
(and thus the finite-size scaling of the heat flux in the nonequilibrum
setup). It turns out that $\delta$ has a dependence on $\sigma$ which is different
from the nonlinear model described above: for instance,  for the momentum-exchange model, superdiffusive transport occurs for $\frac12<\sigma<\frac32$ instead of $0<\sigma<1$.
Moreover, the associated fractional diffusion equation in the hydrodynamic 
limit \cite{suda2021family} has a different order. Thus the two models belong to different classes.

\begin{figure}
\centering
\includegraphics[width=0.7\textwidth]{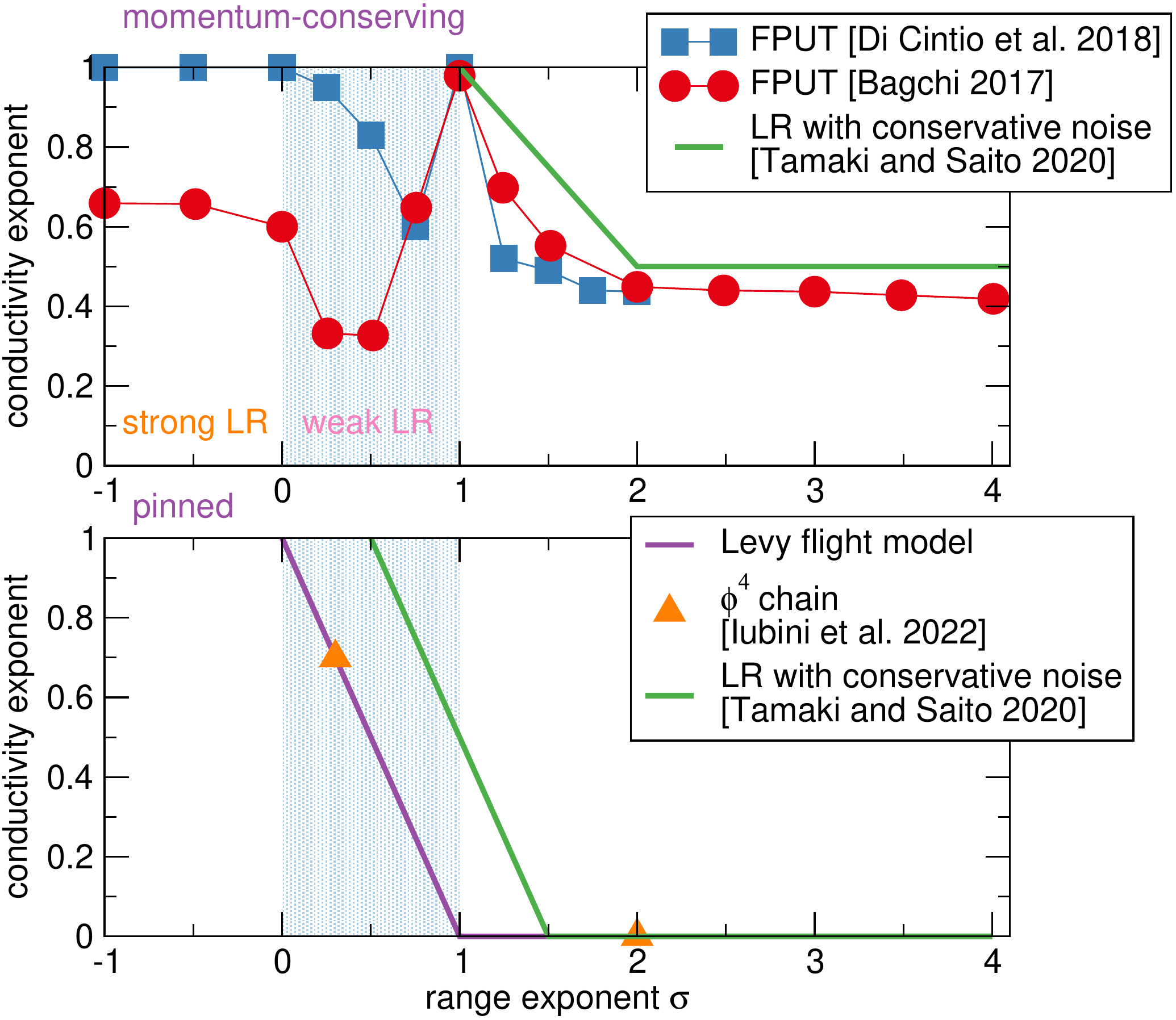}
\caption{Long-range chains: dependence of the thermal conductivity exponent 
$\gamma(\sigma)$ on the exponent controlling the interaction
range $\sigma$. Upper panel: case of models 
momentum-conserving  FPUT (\ref{eq:bagchi}  
and (\ref{eq:olivares} taken from \cite{Bagchi2017} and 
\cite{DiCintio2018}. 
Lower panel: case with pinning potential (\ref{eq:lrphi4})
taken from \cite{iubini2021hydrodynamics}; the solid 
purple line is what predicted from the Levy flight master equation
(\ref{eq:ME}).
In both panels, the solid green lines are the analytical result
for the related momentum-collision model, see \cite{Saito2018}
for details.
} 
\label{fig:explr}
\end{figure}

The situation is even more complicated for models (\ref{eq:bagchi}) and (\ref{eq:olivares}). The numerical results
indicate that the exponent $\gamma$ 
of finite-size conductivity depends in 
a non-trivial way on $\sigma$ \cite{Bagchi2017,DiCintio2018}, 
see Fig. \ref{fig:explr}.
An intriguing feature is also that for $\sigma=1$ 
the conductivity diverges almost linearly with the system size and the temperature profile has a negligible slope \cite{Bagchi2017,DiCintio2018}. This 
ballistic transport regime is unexpected and may perhaps
suggest that the models are close to some  
(yet unknown) integrable limit.
To try to rationalize the results so far available,
in Fig. \ref{fig:explr} we collected the data of the exponent 
$\gamma$ as a function of the range exponent $\sigma$.

At equilibrium, there is also evidence of dynamical 
scaling of the correlation functions of sound modes, akin to the one seen in the short-range case Eq. (\ref{eqscalS}) but  with a $\sigma$-dependent dynamical exponent $z(\sigma)$ \cite{DiCintio2019}. 
Within the accessible time and size ranges, it is also found that the short-range limit is hardly attained, even for relatively large values of $\sigma$. 

In conclusion, there is plenty of evidence that transport in long-range interacting systems may be anomalous, but a systematic understanding of the different universality classes, akin to the short-range case, is lacking.

\subsection{Low-dimensional fluids} 
\label{sec:mpc}

So far we have discussed the case of lattice models, mostly in
one dimension. To test the generality of the results
and their universality, we also discuss the case of 
low-dimensional fluids. Historically, it is in this context that the long-time tails of correlations were first discovered and studied by mode-coupling approximations \cite{Pomeau1975}. 
Although molecular dynamics would be the natural choice, we describe here a different approach, that has been recently used to study anomalous transport. The idea is
to consider effective stochastic processes capable to 
mimic particle interaction through random collisions. A prominent example is the 
Multi-Particle-Collision (MPC) simulation scheme \cite{Kapral1999}, that proved to be very effective 
for the simulation of the mesoscopic dynamics of 
polymers in solution, colloidal and 
complex fluids etc. 

In brief, the MPC method consists in partitioning the system of $N$ particles into $N_c$ cells. The center-of-mass coordinates and velocity
in each cell are computed and  
particle velocities in the cell's center-of-mass frame are rotated around a random axis. The rotation angles are assigned in such a way that the invariant quantities are locally preserved. All particles are then propagated freely, 
for a time interval $\delta t$. 
Physical details of the interactions can be also included. For instance,
energy-dependent collision rates can be considered \cite{DiCintio2017}. 
Interaction with external reservoirs can be implemented by imposing Maxwellian 
distributions of velocity and chemical potentials on the thermostatted cells \cite{Benenti2014}, or via thermal walls at the system boundaries \cite{lepri2021kinetic}.  

For the case of a one-dimensional MPC fluid, since the conservation laws are the same as say, the FPUT model, we expect it to belong to the same KPZ universality class of 
anomalous transport \cite{Narayan02,Spohn2014}. At equilibrium, 
numerical measurements of dynamical 
scaling agrees with Eq. \ref{scaling32} 
both in the strictly 1D 
\cite{DiCintio2015} than in
quasi-one-dimensional case, namely a fluid confined in a
box with a relatively large aspect ratio \cite{DiCintio2017}. Possible dimensional crossovers upon changing the aspect-ratio
are also demonstrated \cite{DiCintio2017}.

\begin{figure}
\centering
\includegraphics[width=0.7\textwidth]{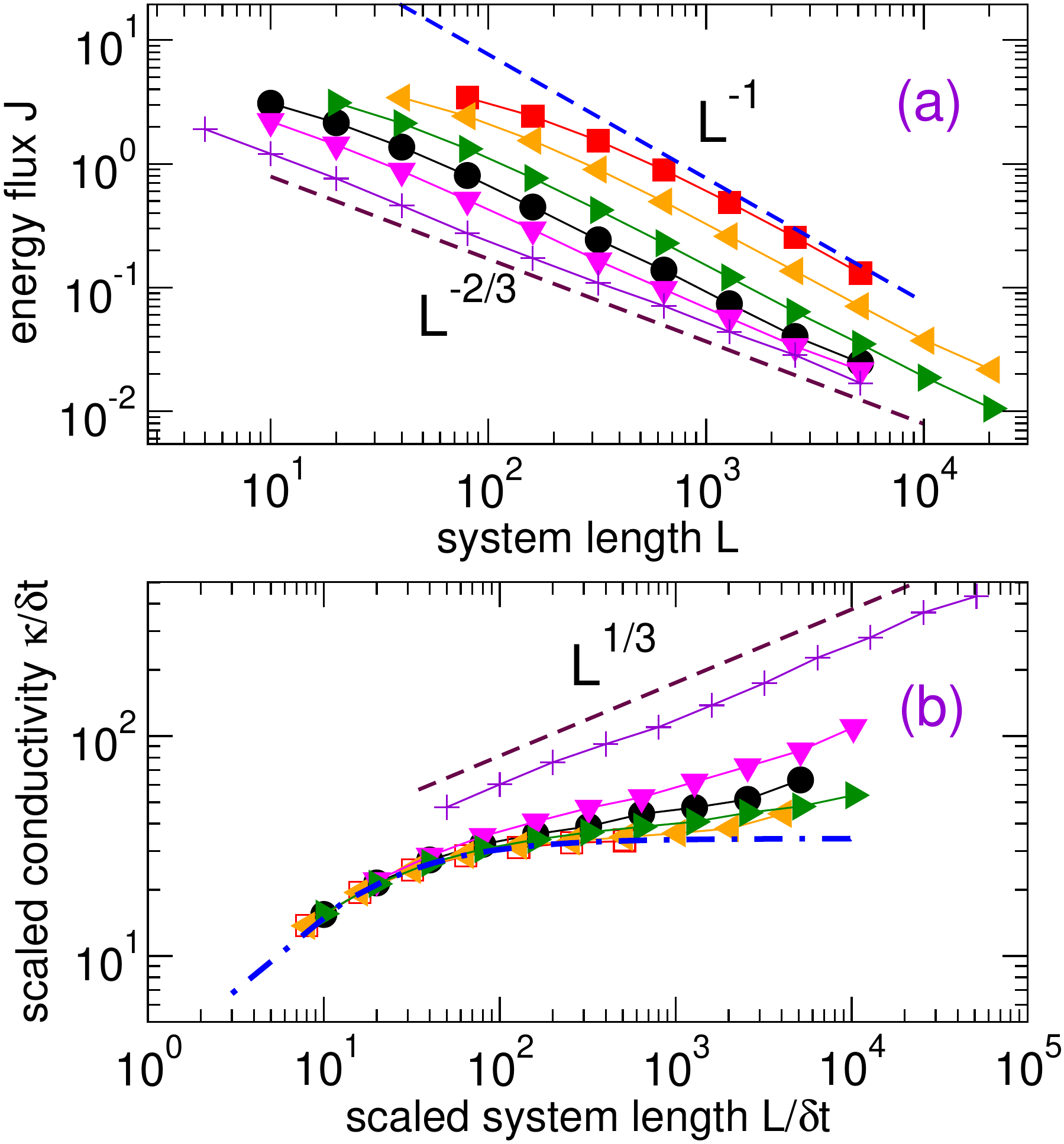}
\caption{MPC fluid confined in a $1d$ box of length $L$.
The MPC collision is performed at regular times steps, separated by a constant time interval of 
duration $\delta t$. Particles interact with two thermal walls at temperatures 
$T_0=4$, $T_L=2$: when a particle crosses the boundary
it is re-injected inside with a new velocity drawn at 
random from a Maxwellian distribution at the 
wall temperature. (a) the energy current $J$ as a function of the box size $L$ for increasing collision times $\delta t=0.1, 0.5, 1.0, 2.0,
5.0$ and $10$ (bottom to top). 
The upper and the lower dashed lines  correspond
to the scaling of normal and anomalous transport, respectively.
(b) The heat conductivity $\kappa=JL/(T_L-T_0)$ with the system size $L$;
the dot-dashed blue line is a fit of the data for $\delta t=10$
with the functional form $34.2x/(12.2+x)$, 
suggested by kinetic theory, see Eq.(\ref{JN}).
} 
\label{fig:fscaling}
\end{figure}

Evidence of superdiffusive heat transport is also 
found in the open setup,
where the 1D MPC fluid interacts with two heat reservoirs
modeled as thermal walls \cite{lepri2021kinetic}.
In Fig. \ref{fig:fscaling} we report
the results of simulations that clearly show 
anomalous transport in the KPZ universality class for the regime of
small enough collision times $\delta t$
(crosses).
Moreover, upon increasing $\delta t$ one observes a 
clear crossover from a normal/kinetic regime to an anomalous/hydrodynamic one, above a characteristic 
size, which can be estimated to
be of order $(\delta t)^{3}$.
This is in agreement with the scenario presented 
above for almost-integrable systems  \cite{Zhao2018,Miron2019,lepri2020too}. 

For the more realistic case of a genuine 3D MPC fluid
for large aspect ratios of the simulation box, a crossover from 3D to one-dimensional (1D) abnormal behavior of the thermal conductivity occurs \cite{luo2021heat}. 
The transition from normal to abnormal transport is 
well accounted for by the decomposition (\ref{Jtot2}) of the energy current, and the three-regimes scenario
described in Section \ref{sec:integ} is again observed in the 
weakly collisional case, where the mean free path is 
large enough (i.e when the frequency of the 
MPC move is small). This confirms that superdiffusive heat transport persists also for almost 1D fluids over a large range of sizes.

To conclude this overview, we register a growing 
interest and evidence for anomalous heat transport
also in other condensed-matter systems. For instance,
the thermal conductivity for one-dimensional electronic fluids has been recently examined \cite{samanta2021thermal}.  It is argued that
at lowest frequencies or longest length scales, the thermal transport is dominated by L\'evy flights of low-momentum bosons that lead to a fractional scaling, 
$\omega ^{-1/3}$ and $L^{1/3}$ 
of heat conductivity with the frequency 
and system size, respectively.

\section{Thermal management and conversion}
\label{sec:manag}
In this Section we describe some application of 
the above concepts to achieve control of energy currents, enhanced energy conversion
and inverse coupled currents, flowing against applied thermodynamic forces.

\subsection{Thermal rectification}

Generally speaking, a thermal rectifier (or thermal diode) is a device that 
allows the heat flow from one end to the other of it, but it inhibits the flow in the opposite direction. 
While the design of a thermal rectifier~\cite{TerraneoCasati2002,li_thermal_2004,Peyrard2006,LiCasati2006,HuZhang2006,YangLi2007,LiLi2012,BenentiPeyrard2016} 
is fully compatible with 
the Fourier law, non-Fourier transport may introduce useful features.

Let us first define a suitable rectification coefficient. We consider the 
heat flow in, say, the $x$ direction, with heat baths imposed at 
the left and right boundaries of a system, 
$T(x=0)=T_1$ and $T(x=L)=T_2$, respectively.
Thermal rectification can occur if there exist some 
features that break the left-right symmetry of the system.
In the case of Fourier transport, we should have a
local thermal conductivity that depends on the position 
$x$ and on the local temperature $T(x)$. 
Using the Fourier law, we can then write
\begin{equation}
  \label{eq:tdistrib}
T(x) = T(0) + \int_0^x d \xi\,\dfrac{J_f}{\kappa[\xi,T(\xi)]}  \, ,
\end{equation}
where $J_f$ if the forward heat flow (from left to right, 
under the condition $T_1>T_2$) and $\kappa$ the local 
thermal conductivity. By solving this equation under the 
boundary condition $T(x=L)=T_2$, we can determine $J_f$. 
If the boundary conditions are reversed, $T(x=0)=T_2$ and $T(x=L)=T_1$,
we obtain another dependence on $x$ of local temperature and local 
thermal conductivity. The resulting backward heat flow 
$J_b$ (from right to left) can be different from the forward heat flow,
as a result of either inhomogeneity in the material or in the 
geometry. The rectification coefficient can then be defined as 
\begin{equation}
f_{r}=\frac{(J_{+}-J_{-})}{J_{-}}\times100\%,
\label{eq:frectification}
\end{equation}
where $J_{+}={\rm max}\{J_f,J_b\}$ and 
$J_{-}={\rm min}\{J_f,J_b\}$.

Summarizing the previous discussion, 
we need two basic ingredients for thermal rectification: 
a temperature-dependent thermal conductivity 
and the breaking of the (left-right) inversion symmetry 
in the direction of the heat flow.
Here, we consider the case of electrical insulators in
which heat is only carried by phonons.
The two basic ingredients can then be found in simple 
models of one-dimensional lattices.
Let us consider the Hamiltonian
\begin{equation}
H=\sum_{i=1}^{N} \left( \frac{p_i^2}{2 m_i}+ V_i(q_i)
\right) +
\frac{1}{2}\sum_{i=1}^{N-1}\left(q_{i+1}-q_i\right)^2,
\label{eq:Hamiltonian_spacer}
\end{equation}
where $q_i$ denotes the displacement from the
equilibrium position of the $i$-th particle with mass $m_{i}$ and
momentum $p_{i}$, and $V_i(q_i)$ is a nonlinear on-site potential.
Thermal rectification requires some inhomogeneity in the system, 
for instance, one could consider segmented chains with different 
nonlinearities in the different parts, or a mass-graded system. 
The nonlinearity is needed to obtain a temperature dependence of
the phonon bands. 
In the presence of a thermal gradient, the effective phonon 
frequencies can depend, for a given position $x$, 
on the orientation of the gradient. 
As a result, one can have either a 
good matching of the phonon bands at the interfaces between 
different parts of the material (say, for forward thermal bias, 
$T_1>T_2$), or a mismatch (for backward bias, $T_1<T_2$). 
In the first case, the thermal conduction is expected to be much 
higher than in the latter.
The discussion so far could be carried within the framework of 
Fourier heat transport,
and indeed theoretical models of nonlinear systems which 
obey the Fourier law have been proposed, for which
a rectification factor up to the order of $10000\%$ has
been found.

Besides experimental difficulties
(phononic devices~\cite{ChangZettl2006,KobayashiTerasaki2009,SawakiTerasaki2011,KobayashiTerasaki2012,Schmotz_2011,RobertsWalker2011,TianLiu2012,WangChen2017,LiuMizuta2021} so far are limited to $f_r\approx 70\%$),
there is a main conceptual limitation. 
For a system described by the Fourier law, rectification 
rapidly decays to zero as the size increases. This effect is
due to the fact that rectification is a nonlinear phenomenon,
which vanishes in the linear response regime.
For a given temperature bias, the temperature gradient
decreases as the system size increases, and therefore 
the linear response regime is approached more and more 
with increasing the system size. 
To address larger system sizes is, on the other hand, a 
practical necessity, since it is difficult to apply large temperature 
biases on small scales.

\begin{figure}[h]%
\centering
\includegraphics[width=1.0\textwidth]{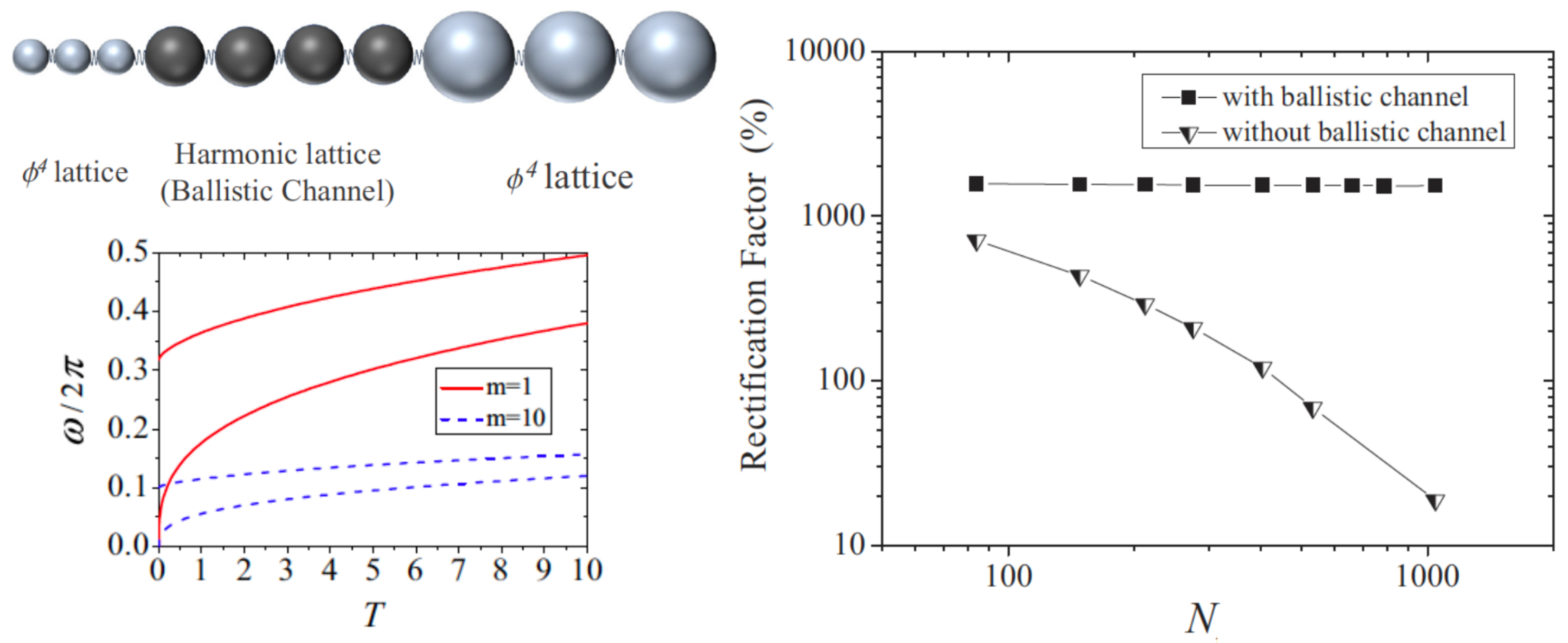}
\caption{Top left: schematic drawing of a thermal rectification
model based on mass-graded leads, connected by a ballistic
channel. Bottom left: boundaries of the 
effective phonon band as a function of 
temperature, for lighter masses ($m=1$, solid red lines)
and for heavier masses ($m=10$, dashed bliue lines).
Right: rectification factor versus 
the overall system size $N$, 
with ballistic channel (squares) and
for the $\phi^4-\phi^4-\phi^4$ model (triangles),
for ${\rm max}(T_L,T_R)=9.5$, ${\rm min}(T_L,T_R)=0.5$,
$N_{L}=N_{R}=10$, $m_{L}=1$, $m_{C}=4.5$, $m_{R}=10$, $\gamma_{L}=\gamma_{R}=1$,
and $\gamma_{C}=0$ ($\gamma_{C}=1$) 
for the model with (without) ballistic channel.
Adapted with permission from~\cite{ChenCasati2018}. Copyright @ 2018 American Physical Society.}
\label{fig:spacer}
\end{figure}

Non-Fourier transport offers a possibility to solve this problem,
by inserting a ballistic channel between the two, anharmonic
and asymmetric, leads~\cite{ChenCasati2018}
(see Fig.~\ref{fig:spacer}, top left, 
for a schematic drawing of the model). 
For instance, one can consider 
the $\phi^4$ lattices, 
$V_i(q_i)=\frac{\gamma_i q_i^4}{4}$ in Eq.~(\ref{eq:Hamiltonian_spacer}). 
The overall system consists of $N_{L}$ ($N_{R}$) particles with
mass $m_{L}$ ($m_{R}$) and strength of the on-site potential 
$\gamma_L$ ($\gamma_R$) in the left (right) lead. 
The two anharmonic leads
are connected by a ballistic channel, that is, by a purely harmonic
central chain of $N_C$ particles with mass $m_C$ and zero 
on-site potential, $\gamma_C=0$. The total system size is $N=N_{L}+N_{C}+N_{R}$.

The asymmetry needed for heat rectification is here provided 
by considering a mass-graded system, with $m_L<m_R$.
The mechanism for rectification can be explained in terms 
of matching/mismatching of the phonon bands for the two leads,
when the higher temperature is on the side of the heavier/lighter
masses. An effective phonon analysis 
predicts the phonon spectrum
in the band $\sqrt{1.23\,T^{2/3}/m}\le \omega \le 
\sqrt{(4+1.23\,T^{2/3})/m}$.
This band is shown in the bottom-left panel 
of Fig.~\ref{fig:spacer}. 
A weak temperature dependence
is observed for heavy masses, since in this case the nonlinearity is 
weak, while the temperature dependence is much stronger 
for light masses.
The phonon bands for the two leads then significantly overlap 
when the higher temperature is applied to the heavier lead,
with band mismatch for the reverse thermal bias configuration.
The ballistic channel is then chosen with phonon band that 
has a significant overlap with the phonon bands of both leads,
as it is the case for mass $m_C$ intermediate between $m_L$ and $m_R$.

The advantage of a non-Fourier, ballistic channel is clear from 
the right panel of Fig.~\ref{fig:spacer}, showing
the size-dependence of the rectification factor $f_r$  when the intermediate channel is either diffusive 
($\gamma_C=\gamma_L=\gamma_R$) or
ballistic ($\gamma_C=0$). While in the former case 
($\phi^4-\phi^4-\phi^4$ model) the rectification 
factor rapidly decays with the system size, in the 
latter case ($\phi^4$-harmonic-$\phi^4$ model)  
the rectification factor is size-independent. 
Such a result is a consequence of the flat temperature profile in 
the ballistic channel. That is, the temperature drop happens only 
in the leads and therefore the temperature gradient there is not affected 
by the length of the ballistic channel. The linear response regime,
which would imply no rectification, is therefore never reached, in 
contrast with the case where the whole system is a $\phi^4$
lattice, which obeys the Fourier law.

\subsection{Coupled Transport}

In this section, we discuss the possibilities 
offered by anomalous heat transport for heat to work conversion. 

We consider steady-state transport of two coupled flows, 
induced by two thermodynamic forces. For concreteness, we 
focus on thermoelectricity~\cite{Shakouri2011,Dubi2011,Benenti2017,Narducci2021}, 
where the two coupled flows are heat and 
particle flow, induced by gradients of 
temperature and electrochemical potential. 
However, the discussion that follows could be easily reformulated 
to other cases, like thermodiffusion, where 
the flow coupled to heat is particle flow.
In thermoelectric devices part of the heat flow is converted into 
useful work. To characterize the performance of a heat engine 
we should consider different features. 

First of all, 
the engine efficiency, defined as the ratio 
$\eta=W/Q_h$ of the output work $W$ 
over the heat $Q_h$ extracted from the hot reservoir. 
The second law of thermodynamics tells us that the efficiency is 
upper bounded by the Carnot efficiency $\eta_C=1-T_c/T_h$, with 
$T_h$ and $T_c$ temperature of the hot and cold reservoir,
coupled to the system (the ``working fluid'') 
where the energy conversion occurs.

Moreover, one should also consider the output power, since an
ideal engine approaching the Carnot efficiency for a quasi-static,
infinitely slow transformation would be of no practical use. 
Indeed, the delivered power would vanish in that limit. 

Finally, also the constancy in the power supplied by 
the engine is important, and power fluctuations are expected to 
play an important role when dealing with nanoscale engines.

Note that, while we focus our discussion on power production, 
one could equally well consider refrigeration, after an appropriate
reformulation of the problem, where the coefficient of performance
(heat extracted from the cold reservoir over the absorbed power)
and the cooling power are considered as key quantities.

Thermoelectric transport can be conveniently formulated within linear 
response. Since linear response is based on the expansion of currents,
to linear order, around local equilibrium conditions, the validity 
of such approach requires that the temperature drop $\Delta T$
and the electrochemical potential drop $\Delta\mu$ on the scale of the 
relaxation length are such that $\Delta T<< T$ and $\Delta\mu<< k_B T$,
where $k_B$ is Boltzmann constant and $T$ the local temperature.
In thermoelectric materials at room temperature, 
electrons are typically thermalized by inelastic 
electron-phonon scattering and the relaxation length is of 
some tens of nanometers. In this case linear response is usually 
a good approximation, even though the temperature differences between 
the two reservoirs can be large. 
For instance, in the proposed application of the automotive industry 
to generate electricity from the waste heat in a vehicle's exhaust pipe,
the hot reservoir (the exhaust gases) is at temperature 
$T_h\approx 600-700$ K, while the cold reservoir (the environment) 
is at room temperature, $T_c\approx 270-300$ K. In spite of the large
temperature difference, linear response can be used, since the 
temperature drop from $T_h$ to $T_c$ takes place on the scale of 
a few millimeters, and therefore on the scale of the relaxation length
(around 10 nanometers) the temperature drop is of the order of 
$0.003$ K, much smaller than the local temperature.

Within linear response, the relation between currents and generalized forces is linear \cite{callen,mazur}. In particular, for thermoelectric transport we have
\begin{eqnarray}
\left\{
\begin{array}{l}
J_e=L_{ee} \mathcal{F}_e + L_{eu} \mathcal{F}_u,
\\
\\
J_u=L_{ue} \mathcal{F}_e + L_{uu} \mathcal{F}_u,
\end{array}
\right.
\label{eq:coupledlinear}
\end{eqnarray}
where $J_e$ is the electric current density, 
$J_u$ is
the energy current density, and the conjugated thermodynamic forces are
$\mathcal{F}_e=-\nabla (\mu/eT)$ and
$\mathcal{F}_u=\nabla(1/T)$, where $\mu$ is the electrochemical potential and $e$ is the electron charge. The coefficients $L_{ab}$ ($a,b=e,u$) are known as kinetic or Onsager coefficients.
Note that the heat current is
$J_h=J_u-(\mu/e) J_e$, namely it is the difference between 
the total energy current $J_u$ and the ``ordered'' part of it, i.e. the
electrochemical potential energy current $(\mu/e) J_e$.

The kinetic coefficients $L_{ab}$ are related to the more familiar thermoelectric transport coefficients: the electrical conductivity $\sigma$, the thermal conductivity $\kappa$, the thermopower (or Seebeck coefficient) $S$, and the Peltier coefficient $\Pi$:
\begin{equation}
\sigma=-e\,\left(\frac{J_e}{\nabla\mu}\right)_{\nabla T=0}=\frac{L_{ee}}{T},
\label{eq:el_conductivity}
\end{equation}
\begin{equation}
\kappa=-\left(\frac{J_h}{\nabla T}\right)_{J_e=0}=
\frac{1}{T^2}\frac{\det {\bm L}}{L_{ee}},
\label{eq:th_conductivity}
\end{equation}
\begin{equation}
S=-\frac{1}{e}\left(\frac{\nabla \mu}{\nabla T}\right)_{J_e=0}=
\frac{1}{T}\left(\frac{L_{eu}}{L_{ee}}-\frac{\mu}{e}\right),
\label{eq:seebeck}
\end{equation}
\begin{equation}
\Pi=\left(\frac{J_h}{J_e}\right)_{\nabla T=0}
=\frac{L_{ue}}{L_{ee}}-\frac{\mu}{e}.
\label{eq:peltier}
\end{equation}
For systems with time reversal symmetry,
due to the Onsager reciprocal relation $L_{eu}=L_{ue}$, 
and therefore $\Pi=TS$.

When the above Onsager relation is valid, 
the thermoelectric performance is governed by the thermoelectric figure of merit
\begin{equation}
ZT=\frac{\sigma S^2}{\kappa}.
\label{eq:ZT-def}
\end{equation}
Thermodynamics imposes only a lower bound on the figure of merit: 
$ZT\ge 0$.
The thermoelectric conversion efficiency is a monotonic increasing function of $ZT$,
with $\eta=0$ at $ZT=0$
and $\eta\to\eta_C$ in the limit $ZT\to\infty$.
Nowadays, most efficient thermoelectric devices operate at around
$ZT \approx 1$, corresponding to a maximum efficiency
about $15\%$ of the Carnot efficiency. 
On the other hand, it is generally accepted that 
$ZT > 3-5$ is the target value for a commercially competing
thermoelectric technology. 
Indeed, reaching these values 
would yield a maximum efficiency about $40\%$
of the Carnot efficiency,
thus making thermoelectric devices on par with 
other widely used heat engines.
It is an elusive challenge to
increase the thermoelectric efficiency,
since the transport coefficients $S,\sigma,\kappa$ are generally interdependent.
For instance, the phenomenological Wiedemann-Franz law
states that
$\sigma$ and the electronic contribution $\kappa_e$
to $\kappa$ are proportional, so that it is 
not possible to independently increase $\sigma$ and decrease 
$\kappa$, as desirable to enhance the figure of merit $ZT$.
It is therefore of great importance to
understand which physical mechanisms 
might allow to independently control
the above transport coefficients, and in particular
to violate the Wiedemann-Franz law.
Note that the thermal conductivity $\kappa=\kappa_e+\kappa_p$ also 
includes the contribution $\kappa_p$ from phonons and photons, 
so that improving the efficiency of 
energy conversion for electrons by itself does not guarantee a high $ZT$.
At the same time, one should be able to manipulate phonon transport 
to reduce the overall thermal conductivity.

In the case of non-integrable, 
momentum-conserving systems, the thermal conductivity
$\kappa$ diverges sub-ballistically
with the system size $L$.
On the other hand, a theoretical argument~\cite{Benenti2013} 
predicts that in 
such systems the electrical conductivity is ballistic, 
that is, $\sigma\propto L$, and the Seebeck saturates 
with the system size. This leads to a blatant violation 
of the Wiedemann-Franz law, since $\sigma/\kappa\to\infty$
in the thermodynamic limit $L\to\infty$. At the same time,
since $S\sim L^0$, also $ZT$ diverges.

\begin{figure}[h]%
\centering
\includegraphics[width=1.0\textwidth]{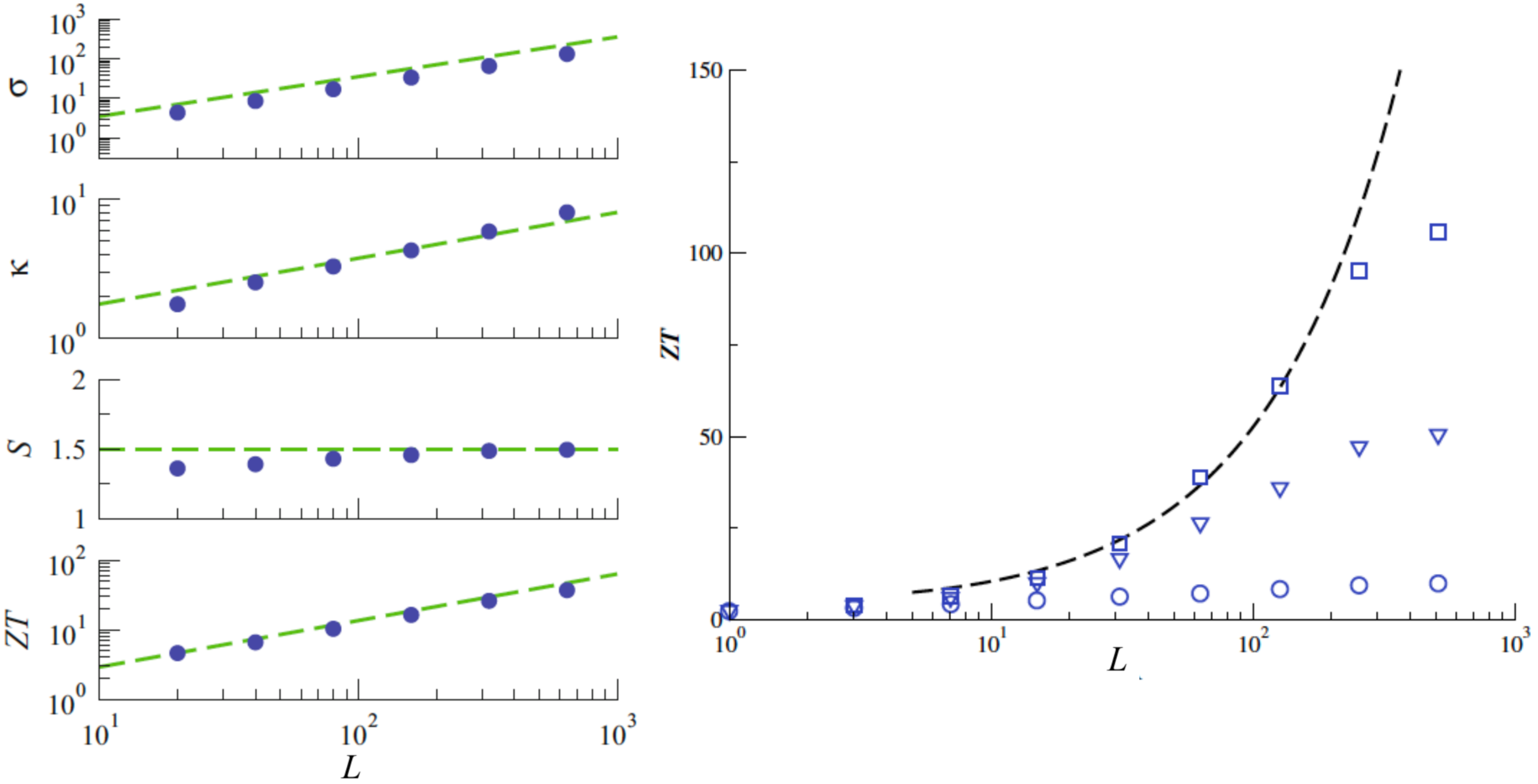}
\caption{Left: Thermoelectric transport coefficients for the 
two-dimensional MPC model as a function of the system size. 
The dashed lines correspond to the theoretical 
predictions $\sigma\sim L$, 
$\kappa\sim\log L$, $S=3/2$, and $ZT\sim L/\log L$.
Right: $ZT$ vs. $L$, with the above theoretical 
expectation (dashed line) compared with simulation where a 
stochastic noise breaking momentum conservation is added, of
strength growing from top to bottom.
Adapted with permission from~\cite{Benenti2014}. 
Copyright @ 2014 Institute of Physics.}
\label{fig:ZT}
\end{figure}

Such prediction has been confirmed in several momentum-conserving models:
in a diatomic chain of hard-point colliding particles
\cite{Benenti2013},
in a two-dimensional system \cite{Benenti2014}, with the dynamics
simulated by the MPC method
discussed in Section \ref{sec:mpc} (see Fig.~\ref{fig:ZT} left),
and in a one-dimensional gas of particles with
screened (nearest neighbour) Coulomb interaction \cite{Chen2015}.
In all these models, collisions are elastic.
On the other hand, when noise breaking momentum conservation is added,
$ZT$ becomes asymptotically independent of the system size, as expected 
for a diffusive transport regime. However, when noise is weak
$ZT$ can grow up to large values before saturating
(see the right panel of Fig.~\ref{fig:ZT}).

Besides efficiency, also power and constancy 
deserve a close inspection.
Indeed, an ideal heat engine should 
operate as close as possible to the Carnot efficiency,
deliver large power and exhibit small fluctuations.
A trade-off between these three desiderata of a heat 
engine can be obtained for a broad class 
of systems on the basis of thermodynamic uncertainty 
relations (TURs) 
(see~\cite{Seifert2019review,Horowitz2020} for perspective 
papers on TURs). More precisely, we refer to 
classical systems described by thermodynamically 
consistent rate equations on a 
discrete set of states or modeled with 
overdamped Langevin dynamics, under steady-state 
conditions and with time-reversal symmetry.
In that case, from the TUR for the ``work current'', namely 
for power, it is possible to show that~\cite{Pietzonka2018}
\begin{equation}
\mathcal{Q}\equiv P\frac{\eta}{\eta_C-\eta}\frac{k_B T_R}{\Delta_P}\le\frac{1}{2},
\label{eq:seifertbound}
\end{equation}
where the (steady-state) power fluctuations are given by
\begin{equation}
\Delta_P\equiv \lim_{t\to\infty}[P(t)-P]^2\,t.
\label{eq:TUR}
\end{equation}
Here $P(t)$ is the mean delivered power up to time $t$.
Note that, since $P(t)$ converges for $t\to\infty$ to 
$P$ as $1/\sqrt{t}$,
an additional factor of $t$ in (\ref{eq:TUR}) 
is needed to obtain a finite limit for $\Delta_P$.
Bound (\ref{eq:seifertbound}) tells us that it is not possible to go 
arbitrarily close to the Carnot efficiency while 
keeping at the same time finite power and non-diverging fluctuations.

Note that non-integrable, momentum-conserving 
systems, for which heat conductivity is anomalous, 
can achieve the bound $\mathcal{Q}=\frac{1}{2}$
when approaching the Carnot efficiency~\cite{Benenti2020}.
On the other hand, for (classical and quantum) systems
described by scattering theory the upper bound
for $\eta\to\eta_C$ is lower, $\mathcal{Q}=\frac{3}{8}$~\cite{Benenti2020}.
Therefore, interactions are necessary to achieve the optimal 
performance of a steady-state heat engine, as it is the 
concretely the case for momentum-conserving systems.

\subsection{Inverse coupled currents}

The usual way to build a thermoelectric heat-engine is
to construct a thermocouple from two thermoelectric materials,
the two ideally having opposite thermoelectric responses.
The reason is that in the Seebeck effect a temperature difference 
pushes charge carriers to the cold side of the material, 
so that a voltage is induced. 
In the case the flow is due to electrons,
as in $n$-type semiconductors, the Seebeck coefficient 
$S$ is negative. On the other hand,
in $p$-type semiconductors, the flow is due to holes, positively charged,
and so $S$ is positive. 
As a consequence, in a thermocouple a circulating current is 
established, and a load attached to the circuit can turn 
the electrical power into some other kind of work (for instance 
the load could be a motor which generates mechanical work).

Momentum-conserving systems offer a conceptually appealing 
possibility to have systems with opposite sign of the Seebeck 
coefficient. A negative Seebeck is here quite counterintuitive 
since it implies the possibility of a particle current against
thermodynamic forces, a possibility, however, not excluded by
thermodynamics for coupled flows. 
For a single flow, a response to an applied static force 
$\mathcal{F}>0$
by generating a current $J<0$ against that force is known as 
absolute negative mobility (ANM).
Such possibility is excluded around a thermal 
equilibrium state~\cite{Cleuren2001,Eichorn2002}, otherwise ANM
could be exploited to construct a perpetuum mobile 
of the second kind, with a single heat bath performing work.
That is to say, the entropy production rate 
$\dot{\mathscr{S}}=J\mathcal{F}$ would be negative, 
against the second law of thermodynamics.
Therefore, ANM can appear only in nonequilibrium setups,
for instance in relation to particle separation~\cite{Reguera2012,Slapik2019}, 
self-propulsion~\cite{Ghosh2014}, 
tracer dynamics in a laminar flow~\cite{Sarracino2016}, 
and also experimentally in semiconductor superlattices~\cite{Keay1995}, 
micro-fluidic systems~\cite{Ros2005}, 
and Josephson junctions~\cite{Nagel2008}.

Conversely, the above limitation does not apply 
for coupled flows, and it is indeed 
possible to have inverse currents in coupled 
transport~\cite{Cividini2018,Wang2020} (ICC).
Referring again for concreteness to thermoelectricity,
the positivity of the entropy production rate 
$\dot{\mathscr{S}}=J_e\mathcal{F}_e+J_u\mathcal{F}_u$
($\mathcal{F}_e,\mathcal{F}_u>0$) can be fulfilled 
even though one of the two induced currents has sign 
opposite to both forces (say, $J_e > 0$ and $J_u < 0$).  
Within linear response, one needs negative Onsager 
cross-coefficients~\cite{Iubini2012,Cividini2018,Wang2020}, a possibility not excluded by thermodynamics.
Indeed, since $J_e =L_{ee} \mathcal{F}_e + L_{eu} \mathcal{F}_u$,
and from the positivity of entropy production $L_{ee}\ge 0$,
it is necessary that $L_{eu}<0$. 

\begin{figure}[h]%
\centering
\includegraphics[width=1.0\textwidth]{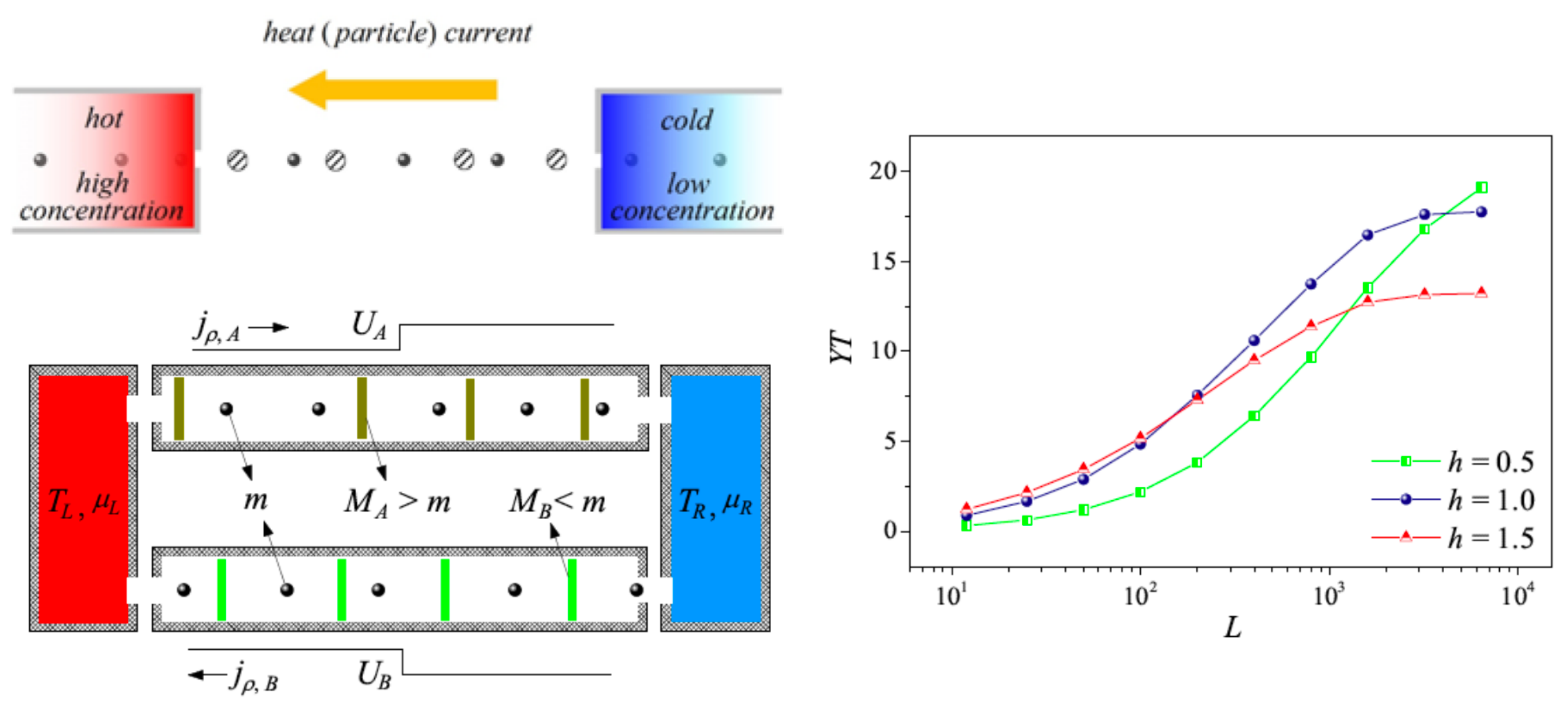}
\caption{Top left: Schematic drawing of ICC, where 
either particle or heat flow is against both the temperature and the concentration gradient.
Bottom left: Sketch of an ICC-based heat engine,
consisting of two channels, between two reservoirs at different temperatures $T_L$ and $T_R$ 
(chemical potentials self-adapt to values 
$\mu_L$ and $\mu_R$ in the steady-state regime). 
For visualization purposes, the two species of charged particles 
in each channel are represented by bullets and rods, respectively. 
A circular current of bullets can form and work can be extracted 
by the applied potentials $U_A$ and $U_B$.
Right: dependence of the figure of merit $YT$ on the 
system size $L=L_A =L_B$ for barrier height
$h=h_A =h_B =0.5,1.0$, and $1.5$.
Adapted with permission from~\cite{Benenti2022}.}
\label{fig:ICC}
\end{figure}

We stress that the ICC phenomenon should not be confused with 
standard thermoelectric transport, where the two thermodynamic 
forces have opposite sign instead. There, 
the motion of particles against 
an electrochemical potential difference is possible thanks to a temperature difference.
In ICC an inverse particle current is obtained, very 
counterintuitively, when 
particles move against both the temperature and 
the concentration gradient (see the top left panel of 
Fig.~\ref{fig:ICC}).

Momentum-conserving systems offer an appealing possibility 
to realize ICC in a Hamiltonian system, 
which can be seen as a classical version of the Lieb-Liniger
model~\cite{Lieb1963,Lieb1963b}:
\begin{equation}
H=\sum_{i} \frac{p_i^2}{2m_{i}}+\sum_{i<j}V(x_{i}-x_{j}),
\label{eq:LL}
\end{equation}
where $m_i\in\{m,M\}$, $V(x)=h$ for $x \le |r|$
and $V(x)=0$ otherwise, with $h\ge 0$ being the 
height of the potential barrier (hereafter we set $r=0$).
Note that for $h=0$, all particles move freely and the system is integrable, while in the other limiting case 
$h\to\infty$, the system reduces to the nonintegrable 
two-species hard-core gas, a paradigmatic model in 
the study of anomalous one-dimensional heat transport.

We can understand the fact that the cross-coefficient $L_{eu}$
is negative as follows.
If we set $\mathcal{F}_e=0$ and $\mathcal{F}_u>0$, 
then the probability for two particles to cross each other 
is higher when the light particle is closer to the hot reservoir
and the heavy particle is closer to the cold reservoir. In this case, 
the relative velocity of the two particles is on average higher 
than in the opposite configuration.
Hence, it is more probable that the relative velocity of the 
two particles injected by the baths is sufficient to overcome 
the potential barrier. 
This creates an unbalance in the particle density for the two species. 
As the temperature difference between the two reservoirs is 
increased, this mechanism can even lead 
to a phase separation~\cite{Garriga2002,WangCasati2017}. 

Thanks to the ICC effect, it is possible to build a 
two-channel heat engine~\cite{Benenti2022} where, as in a thermocouple,
a stationary circular motion of particles is established and
maintained by an applied temperature bias
(see Fig.~\ref{fig:ICC} bottom left). 
In the channels the particles of mass $M_A>m$ (in channel A) and 
$M_B<m$ (in channel B), represented for visualization purposes 
as rods, are reflected back when they hit a channel boundary,
with a newly assigned velocity sampled from 
the equilibrium, thermal distribution 
determined by the temperature of the reservoir
in contact with that boundary.
On the other hand, particles of mass $m$ (depicted as bullets) 
can enter the reservoirs, which in turn 
inject bullets into the channels with rates and energy distribution determined by their temperatures and densities.
Thanks to this effect, it is possible to make the bullets flow
inversely in one channel, say, channel B, by setting $M_B<m$,
so that a clockwise-circulating bullet current through the whole 
system is created.
Work can then be extracted by applying two external bias voltages
$U_A$ and $U_B$ on the two channels, respectively 
(for instance, we could use 
a portion of the kinetic energy of the circulating particles to set to rotation lifting wheels). 
The net effect is that the engine converts a portion of the heat 
flowing from the hot to the cold reservoir into work, 
in a very efficient way. Indeed, numerical simulations 
and a linear response analysis suggest that the Carnot efficiency
can be achieved, under suitable conditions, in the thermodynamic limit. 

Note that within linear response engine efficiency is a monotonously 
growing function of the 
figure of merit $YT$~\cite{Horvat2009,WangLai2009}:
\begin{equation}
YT=\frac{(\sigma_A/L_A)(\sigma_B/L_B)(S_A-S_B)^2}
{(\sigma_A/L_A+\sigma_B/L_B)
(\kappa_A/L_A+\kappa_B/L_B)}\,T,
\label{eq:YT}
\end{equation}
where $\sigma_i$, $\kappa_i$, $S_i$, and $L_i$ stand for 
electrical conductivity, thermal conductivity, 
thermopower, and length of channel $i$ ($i=A,B$). 
Thermodynamics requires $YT\ge 0$, with the efficiency of 
heat to work conversion vanishing for $YT=0$ and achieving the 
Carnot efficiency for $YT\to\infty$. 
Numerical simulations suggest that the Carnot limit can be achieved 
for  $L=L_A=L_B$, $h_A=h_B\sim 1/\sqrt{L}$,in the thermodynamic limit 
$L\to\infty$, thanks to the interplay between ballistic 
particle transport and anomalous, sub-ballistic heat transport,
and to the opposite sign of the Seebeck coefficients in the 
two channels, with the negative Seebeck coefficient 
in the channel with ICC. 

\section{Atomistic simulations}
\label{sec:atom}

Since the 1990s, the developments in the fabrication and characterization of nanostructures have provided the opportunity to test experimentally the predictions of statistical physics models. 
Suitable one-dimensional systems to probe anomalous heat transport are nanotubes, nanowires, and polymer fibers.
Graphene, hexagonal boron nitride, exfoliated transition metal dichalcogenide monolayers, and thin semiconductor membranes have been considered to study thermal transport in two-dimensional systems, as well as possible platform to implement thermal rectification concepts~\cite{LiLi2012}.  

However, it is difficult to reproduce the physics of ideal models in experiments, the direct interpretation of which is often arduous. 
Atomistic simulations provide a much-needed bridge between models and experiments. 
These simulations can entail the specific properties of actual materials and realistic features, such as defects, surface features, and contacts while allowing a detailed insight into the microscopic mechanisms of thermal transport.
In fact, atomistic modeling has been successfully employed both to predict the behavior of nanostructures and to interpret experiments.

In this section, we review the contributions of atomistic simulations to the understanding of the aspects non-Fourier heat transport discussed in the previous sections. After some methodological considerations, we address the diverging thermal conductivity in one- and two-dimensional materials, and thermal rectification. 
Rather than trying to list all the systems that have been studied by molecular simulations, we focus on those closer to statistical models and most likely to exhibit anomalous properties. Further information about simulations of heat transport in nanostructures may be found in~\cite{lepri_simulation_2016}.

\subsection{Molecular simulation methods}

Atomistic methods to simulate thermal transport can be classified into two categories: molecular dynamics (MD) and lattice dynamics (LD). 
Transport coefficients, such as thermal conductivity, can be computed by MD either from the fluctuation of currents at equilibrium (EMD) or, directly, from non-equilibrium (NEMD) simulations. 
In the former case, the thermal conductivity of a system of volume $V$ is obtained as the infinite time limit of the Green-Kubo integral of the heat current autocorrelation function: 
\begin{equation}
    \kappa_{\alpha\beta} = \frac{1}{V k_BT^2}\int_0^\infty dt \langle J_\alpha(t)J_\beta(0)\rangle 
\label{eq:GK}
\end{equation}
where $\alpha,\beta$ are the Cartesian components of the heat current vector $J$.
EMD applies to periodically replicated systems in the bulk limit. 
Non-Fourier transport is characterized by a slow decay of $\langle J(t)J(0)\rangle \propto t^{-(1-\delta)}$, i.e. $\delta\ge 0$ (see Section~\ref{sec:nonequi}).
Hence, proving non-Fourier thermal transport by EMD involves probing the long-time tail of correlation functions, which are affected by large statistical uncertainties. This problem is further aggravated by the poor ergodicity of most low-dimensional systems. 
This issue may be circumvented by running several statistically independent replicas of the same system. 

NEMD and reverse-NEMD~\cite{muller-plathe_simple_1997,kuang_gentler_2010} methods are similar to actual experiments: a system of finite length in the transport direction is connected to two thermal baths at different temperatures. The evolution of the system is simulated until the energy flux ($J$) from the hot to the cold reservoir becomes stationary. When transport is diffusive at stationary conditions $\kappa$ may be estimated directly from Fourier's law as $\kappa=-J/\nabla T$ by estimating $\nabla T$ from the temperature profile. 
In fact, as one normally simulates systems from several nanometers up to at most 10~$\mu$m, the temperature profile is often nonlinear. In these cases, one should calculate the Kapitza conductance of finite-length ($L$) systems as $G(L)=J/\Delta T$, where $\Delta T$ is the temperature difference between the two thermal baths, and estimate the length-dependent thermal conductivity as $\kappa(L)=G\cdot L$~\cite{li_influence_2019}.
The results of finite-size simulations cannot be easily extrapolated to the infinite size limit so proving the divergence of $\kappa$ of realistic materials through NEMD is a deceptive task~\cite{sellan_size_2010}.
In turn, as a non-perturbative method, NEMD is ideal to probe non-linear effects, such as thermal rectification.  
There is a variety of algorithms, {\it thermostats}, to control either the temperature of the thermal reservoirs in direct NEMD or the heat current in reverse NEMD. Whereas in the diffusive transport regime in three-dimensional materials different approaches to NEMD are generally equivalent, the choice of the algorithm becomes critical in low-dimensional systems, especially concerning ergodicity~\cite{li_influence_2019,hu_unification_2020}.

An efficient alternative way of computing thermal transport in the MD framework is the so-called approach-to-equilibrium MD (AEMD)~\cite{lampin_thermal_2013}. AEMD consists of preparing the system out of equilibrium with two regions at different temperatures and monitoring the relaxation to equilibrium. In diffusive conditions, the temperature difference between the hot and cold regions should decay exponentially following the heat equation, and the thermal diffusivity is proportional to the relaxation time. Size effects in this approach are similar to NEMD. 

Taking care of the above-mentioned technical precautions, EMD and NEMD approaches are complementary and provide a versatile framework to perform numerical thermal transport experiments. However, the classical nature of MD limits the direct comparison to experiments when quantum effects matter, i.e. when the temperature is lower than the Debye temperature, $\Theta_D$ of the system. For example, $\Theta_D > 1500$ K for carbon-based materials and $\sim600$ K for silicon. 
While quantum corrections can be implemented in NEMD simulations with spectral decomposition~\cite{fan_homogeneous_2019}, lattice dynamics and the Boltzmann transport equation (LD-BTE) offer a more natural route to compute quantum phonon transport in solids~\cite{Peierls, omini_beyond_1996, cepellotti_thermal_2016}. 
This approach consists of treating the normal modes of a solid as a gas of quantum quasi-particles that diffuse according to the Boltzmann equation at stationary conditions:
\begin{equation}
\vec{v}_i(q)\cdot \nabla T \frac{\partial n_i(q)}{\partial T} = \left[\frac{\partial n(q)}{\partial t}\right]_{scattering} 
\label{eq:BTE}
\end{equation}
where $n_i(q)$ and $\vec{v}_i(q)$ are the populations and group velocities of mode $i$ at wavevector $q$ and $T$ is the temperature.
The right-hand side of Eq.~(\ref{eq:BTE}) can be expressed in terms of a  tensor, $\Gamma$, that encompasses both extrinsic, e.g. isotope and boundary scattering, and intrinsic scattering multi-phonon scattering processes.
The latter are usually approximated considering three-phonon processes only, but recent works have highlighted the impact of higher-order four-phonon processes, especially in high thermal conductivity materials~\cite{feng_quantum_2016, feng_four-phonon_2017}.
The solution of Eq.~(\ref{eq:BTE}) gives an expression for the thermal conductivity 
\begin{equation}
\kappa_{\alpha\beta} = \frac{1}{N_q V}\sum_{i,q} c_{i,q} v^{(\alpha)}_i(q)\lambda^{(\beta)}_i(q)
\end{equation}
in terms of the heat capacity $c_i(q)$, group velocity $v_i(q)$ and mean free path $\lambda^{(\beta)}_i(q)$ of the renormalized phonon modes~\cite{barbalinardo_efficient_2020}, also called {\it relaxons}~\cite{cepellotti_phonon_2015}  
A major advantage of LD-BTE is that it offers a direct evaluation of the contribution of each mode to the thermal conductivity and it easily allows one to identify which phonon may lead to anomalous heat transport in the extended size limit. Yet, atomistic LD-BTE is rarely used to probe anomalous transport, and the comparison with MD, potentially very insightful, is often challenging~\cite{turney_predicting_2009}.

\subsection{Anomalous transport in one-dimensional systems}

\begin{figure}[h]%
\centering
\includegraphics[width=1.0\textwidth]{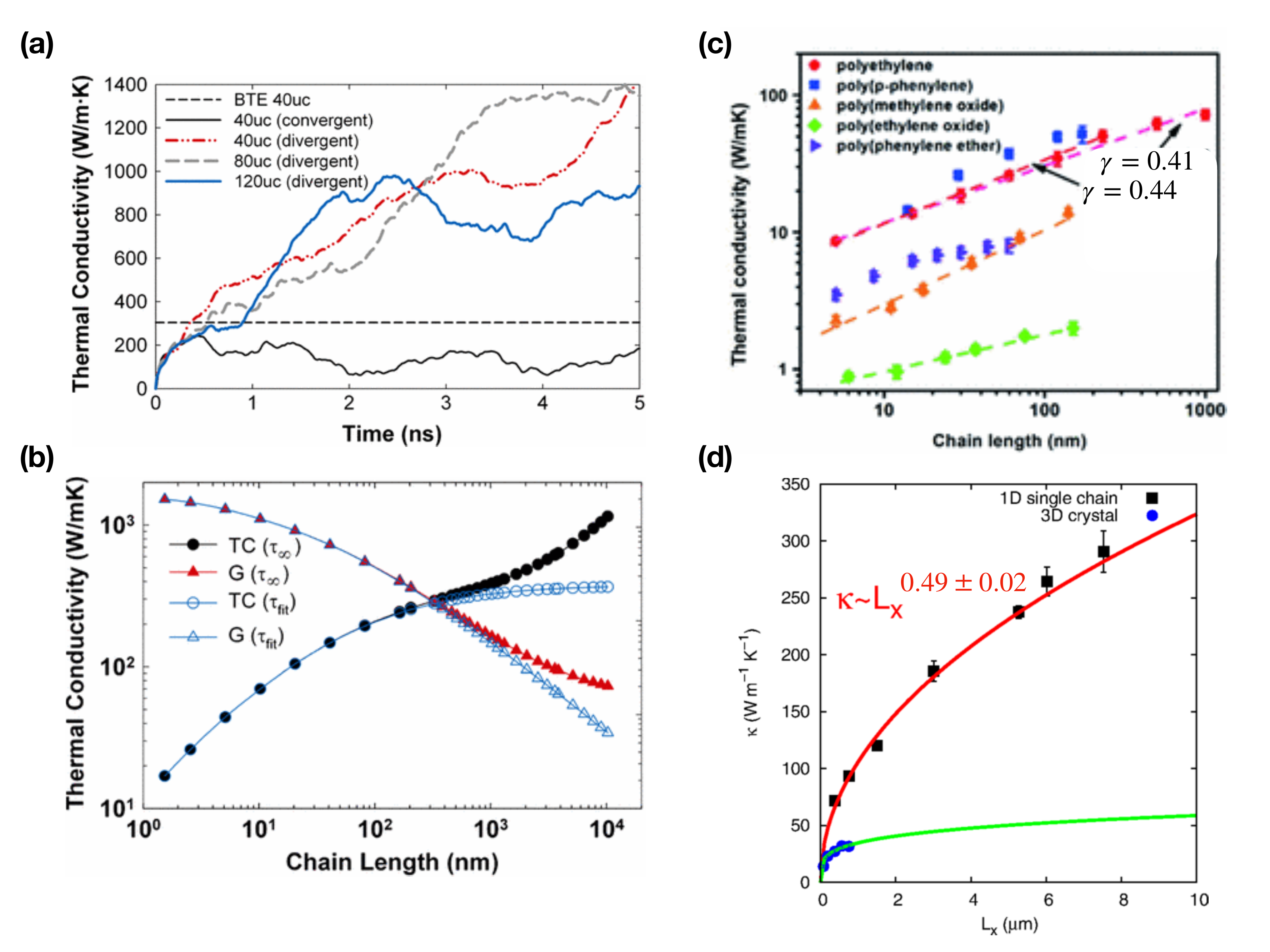}
\caption{Molecular dynamics simulations of isolated polymer chains: (a) Green-Kubo running integral of polyethylene (PE) with a different number of units (uc); (b) extrapolated thermal conductivity (TC) and thermal conductance (G) of PE chain adapted with permission from~\cite{Henry2008}; (c) NEMD results for various polymer chains between 1~nm and 1~$\mu$m-long exhibiting  critical exponents $\gamma\sim 0.4$ adapted with permission from~\cite{liu_length-dependent_2012}; (d) diverging $\kappa$ for PEDOT with $\gamma\sim 0.5$, adapted with permission from~\cite{crnjar_assessing_2018}.}
\label{fig:polymers}
\end{figure}

\paragraph{Isolated polymer chains} 
Isolated polymer chains are the systems that most closely resemble one-dimensional models, such as the FPUT model. While in their bulk form polymer plastics have low thermal conductivity ($\kappa < 0.5$ Wm$^{-1}$K$^{-1}$ for polyethylene (PE) at room temperature), early simulations suggested that single polymer chains can have substantially higher conductivity~\cite{freeman_thermal_1987}.
Realistic EMD simulations predicted a very high thermal conductivity, above 100  Wm$^{-1}$K$^{-1}$ for a single PE chain longer than 10 nm, and possibly diverging $\kappa(L)$ in the infinite-length limit~\cite{Henry2008}. 
The raw data from these simulations, reported in Figure~\ref{fig:polymers}a, show that in different runs of the same system the integral in Eq.~(\ref{eq:GK}) appears to either diverge or converge depending on the initial conditions.  
The thermal conductivity of the infinite PE chain would diverge with length under the assumption that all the four acoustic modes would have diverging mean free path. In turn, $\kappa(L)$ would saturate if one uses a fitted finite value for the relaxation time of these modes, as obtained from the runs in which the Green-Kubo integral converges. 
Further simulations show that increasing the dimensionality of the system to 2D (thin films) and 3D (bulk) suppresses the anomalous transport behavior~\cite{Henry_1D3D_2010}. 
Similarly, at the single molecule level, finite-length fully stretched alkane chains exhibit ballistic transport up to several tens of units~\cite{segal_thermal_2003}, but their conductance is reduced introducing kinks by bending~\cite{li_mechanical_2015}.

NEMD simulations of 1D polymers of several different kinds consistently report $\kappa\propto L^\gamma$ divergence for length up to 1~$\mu$m, with $\gamma \gtrsim 0.4$~\cite{liu_length-dependent_2012}, in good agreement with kinetic theory predictions for the FPUT-$\beta$ model~\cite{Pereverzev2003,Nickel07,Lukkarinen2008,dematteis2020coexistence}. 
Crnjar {\it et al.} investigated anomalous heat transport in poly-3,4ethylenedioxythiophene (PEDOT) 1D chain by  AEMD~\cite{crnjar_assessing_2018}. 
These simulations show a power-law divergence of $\kappa(L)$ with exponent $\gamma\sim 0.5$, possibly corresponding to a new universality class. 
The PEDOT models are up to 7.5~$\mu$m long, about one order of magnitude longer than previous in NEMD numerical experiments. It is nevertheless not proven that such length is sufficient obtain a well-converged estimate of the critical exponents, as shown by systematic studies of size effects in model systems~\cite{Lepri03}.
EMD simulations of PEDOT show similar behavior as for PE: for some trajectories the Green-Kubo integral (see Eq.~(\ref{eq:GK})) reaches a constant value, while for others it diverges.
In general, these works suggest anomalous thermal transport in 1D polymer chains with critical exponents larger than $1/3$.  While it is unlikely that measurements will confirm or refute this hypothesis, these simulations have suggested an effective approach to obtain high-thermal conductivity polymeric materials by drawing PE fibers, so that the PE chains are aligned and stretched as in the ideal models~\cite{shen_polyethylene_2010}.

\paragraph{Carbon nanotubes}
Readily available and much easier to manipulate than polymer chains, carbon nanotubes (CNT) have been among experimentalists' favorite systems to test anomalous heat transport. 
After the first measurements reporting extremely high $\kappa$ in CNTs~\cite{hone_thermal_1999-1,kim_thermal_2001,pop_thermal_2006}, several studies tried to assess the thermal conductivity of carbon nanotubes using either MD or LD simulations~\cite{berber_unusually_2000, moreland_disparate_2004, mingo_length_2005, yao_thermal_2005,zhang_thermal_2005, shiomi_non-fourier_2006, lukes_thermal_2007, donadio_thermal_2007, alaghemandi_thermal_2009, savin_thermal_2009, thomas_thermal_2010, cao_size_2012, saaskilahti_frequency-dependent_2015, ray2019heat}.
Early works confirm the high thermal conductivity observed in experiments. Some of them explicitly address the possibility of anomalous transport, but  they fundamentally disagree on the convergence of $\kappa$ with length. As shown in  Table~\ref{tab:CNT}, most EMD simulations suggest that the thermal conductivity should be finite, based on the convergence of the Green-Kubo integral. In turn, most NEMD simulations show that $\kappa$ keeps growing up to lengths of several $\mu$m. 
As in the case of polymer chains, this uncertain scenario results from the difficulty of converging MD simulations of nonergodic low-dimensional systems, and the sensitivity of the results on specific methodological choices~\cite{salaway_molecular_2014}. 
In any case, both NEMD and AEMD results show anomalous transport - between ballistic and diffusive - in CNTs up to several microns in length.
Ray and Limmer~\cite{ray2019heat} showed that the time evolution of temperature perturbations (as in AEMD) and temperature profiles in NEMD agree with those predicted by a Levy walk model, which is non-diffusive, and the thermal conductivity of CNTs with $L<2~\mu$m diverges as $\kappa\propto L^{1/2}$. This result is supported by the analysis of heat flux fluctuations at equilibrium in finite-length systems, suggesting that $\kappa$ divergence is not produced by large temperature gradients. 

\begin{table}[h]
\begin{center}
\caption{Summary of the MD simulations addressing anomalous heat conduction in CNTs}\label{tab:CNT}%
\begin{tabular}{@{}llll@{}}
\toprule
Reference  & Method  &  $L_{max}$ &  Results \\
\midrule
Zhang\cite{zhang_thermal_2005} & NEMD & 100 nm  &  $\kappa$ diverges \\
Alaghemandi\cite{alaghemandi_thermal_2009} & NEMD & 400 nm & $\kappa$ diverges \\
Savin\cite{savin_thermal_2009} &  NEMD & 500 nm & $\kappa$ diverges \\
Thomas\cite{thomas_thermal_2010} & NEMD & 1.4 $\mu$m & $\kappa$ converges \\
Cao\cite{cao_size_2012} & NEMD & 2.4 $\mu$m & $\kappa$ diverges \\
S\"a\"askilahti\cite{saaskilahti_frequency-dependent_2015} & NEMD & 3 $\mu$m & $\kappa$ diverges \\
Shiomi\cite{shiomi_non-fourier_2006} & AEMD & 25 nm & Non-Fourier propagation \\
Ray\cite{ray2019heat} & AEMD+NEMD & 2 $\mu$m & $\kappa$ diverges \\
Yao\cite{yao_thermal_2005}     & EMD  & 80 nm  & GK integral converges $\kappa_\infty$ size dependent \\
Donadio\cite{donadio_thermal_2007}   & EMD   & 200 nm  & GK integral converges, finite $\kappa_\infty$  \\
Savin\cite{savin_thermal_2009} & EMD & 2000 periods  & GK integral diverges  \\
Pereira\cite{pereira_thermal_2013} & EMD & 108 nm & GK integral converges, finite $\kappa_\infty$  \\
Fan\cite{fan_force_2015} & EMD & 108 nm  & GK integral converges, finite $\kappa_\infty$ \\
\botrule
\end{tabular}
\end{center}
\end{table}

The picture emerging from LD-BTE works is also inconclusive: early calculations showed convergent $\kappa$ if higher-order anharmonic processes are explicitly accounted for~\cite{mingo_length_2005}. However, the same authors eventually suggested that $\kappa$ may converge only when a cutoff to the propagation of low-frequency phonon modes is imposed~\cite{lindsay_lattice_2009}.
These calculations, however, were carried out using a self-consistent solution of the BTE, which has convergence issues when the scattering tensor in the right-hand side of Eq.~(\ref{eq:BTE}) is non-diagonally dominant~\cite{cepellotti_thermal_2016}.
From  these works, it is also clear that solving the LD-BTE in the relaxation time approximation (RTA) does not give the correct physical picture of heat transport in either CNTs or any other low-dimensional system. 

Exploring the dependence of $\kappa$ on the CNT diameter, $d$, both MD and LD-BTE calculations showed that, for a given length, $\kappa$ is higher for thinner CNTs~\cite{cao_size_2012, lindsay_diameter_2010, bruns_heat_2020}. 
As CNTs are not strictly speaking 1D systems, one may argue that the smaller their diameter the closer to an actual 1D system. 
Ref.~\cite{lindsay_diameter_2010} showed that this trend is non-monotonic, as $\kappa$ reaches a minimum for $d\sim 2$ nm, and then slowly rises toward the limit thermal conductivity of graphene. 

\begin{figure}[h]%
\centering
\includegraphics[width=0.95\textwidth]{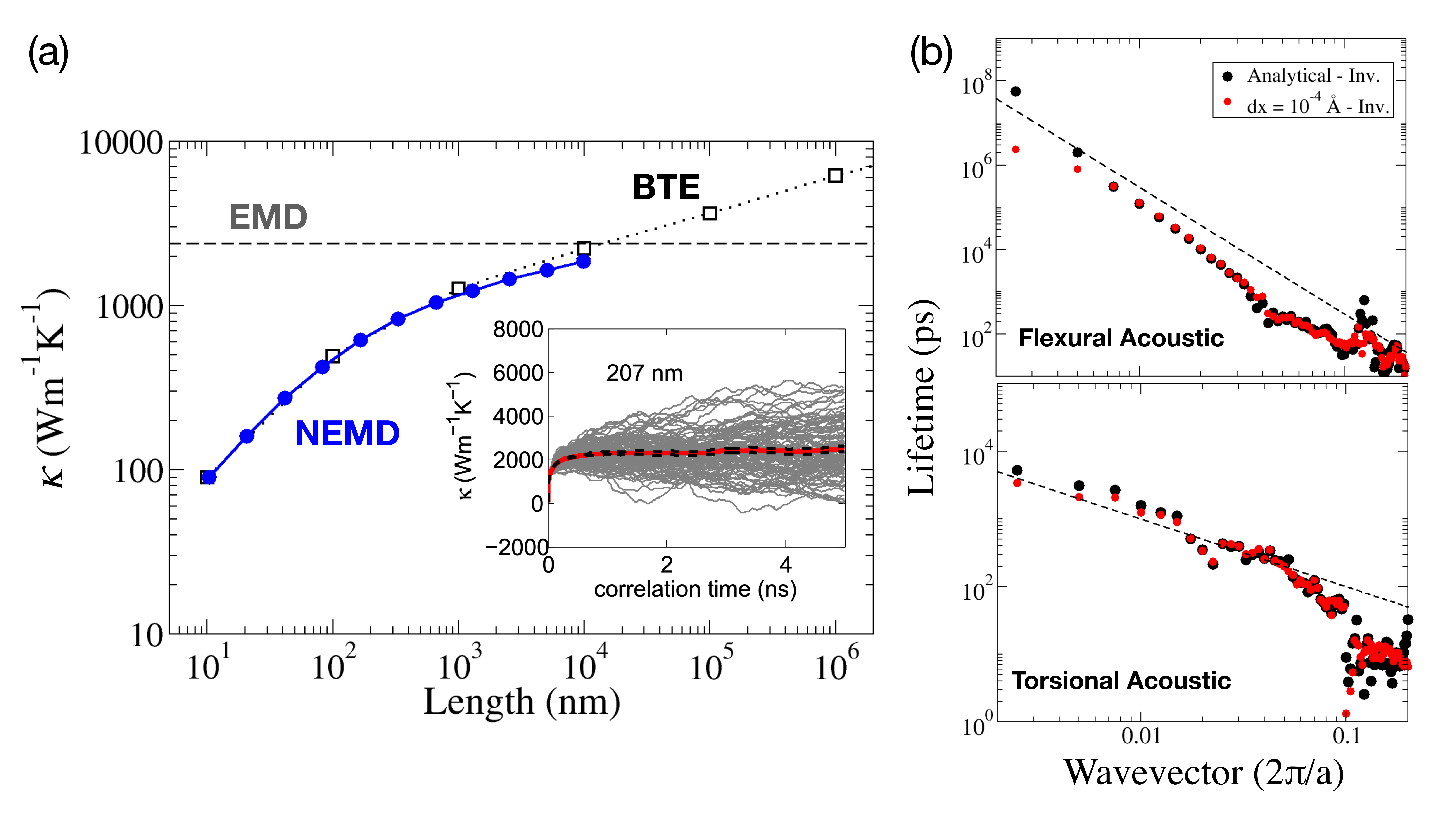}
\caption{(a) Thermal conductivity of a (10,0) CNT from the Boltzmann transport equation (BTE) and equilibrium and nonequilibrium molecular dynamics (EMD, NEMD) simulations as a function of length. The insert shows the convergence of the Green-Kubo integral from an ensemble of EMD runs~\cite{barbalinardo_ultrahigh_2021} (b) Lifetimes of the flexural and torsional acoustic phonon modes of the same CNT from BTE and either analytical or numerical interatomic force constants. The dashed lines indicate the power law for diverging contribution to $\kappa$ in the limit of vanishing wavevector, adapted with permission from~\cite{bruns_comment_2022,barbalinardo_barbalinardo_2022}.}
\label{fig:CNT}
\end{figure}

While the estimates of CNTs $\kappa$ by numerical simulations are in the same ballpark of thousands Wm$^{-1}$K$^{-1}$ at room temperature, there is seldom any quantitative agreement among MD and LD-BTE calculations.   
A couple of recent works have addressed this issue using consistent systems and simulation control parameters for both MD and LD-BTE~\cite{bruns_heat_2020,barbalinardo_ultrahigh_2021}.
These LD-BTE calculations suggest that transport is non-diffusive up to lengths of the order of $\sim$1 mm, but $\kappa (L)$ would probably saturate to a finite value. This determination is however based on the calculation of phonon mean free paths at extremely short wavenumbers, which is very sensitive to numerical errors~\cite{bruns_comment_2022,barbalinardo_barbalinardo_2022}. 
To date, BTE cannot be considered conclusive as for the value of the infinite-length limit thermal conductivity of CNTs. 
Figure~\ref{fig:CNT}b highlights the numerical sensitivity of BTE calculations in the long wavelength limit. 
Bruns {\sl et al.} also highlighted discrepancies between NEMD and LD-BTE~\cite{bruns_heat_2020}.
The main source of these discrepancies is the use of quantum statistics in BTE, while NEMD is Newtonian. If classical statistics is used in the BTE~\cite{barbalinardo_efficient_2020} and the boundary conditions for the finite-length systems are applied correctly~\cite{maassen_steady-state_2015}, the NEMD and BTE agree well for nanotubes up to 10 $\mu$m long.
Furthermore, EMD simulations converge to a value that is higher than the highest NEMD estimate of $\kappa$ for $L \sim ~10\ \mu$m, which makes the two methods substantially agree (Figure~\ref{fig:CNT}a)~\cite{barbalinardo_ultrahigh_2021}.

On the one hand, the analysis of phonon lifetimes from modal analysis of MD trajectories and BTE calculations indicates that the phonons that may lead to anomalous transport are the two flexural acoustic modes (Figure~\ref{fig:CNT}b). 
On the other hand, if the atomic coordinates perpendicular to the CNT axis are constrained, the system is left with only the longitudinal acoustic mode and its thermal conductivity is predicted to diverge as $\kappa\propto L^\alpha$, with $\alpha\sim 0.45$~\cite{barbalinardo_ultrahigh_2021}.
A similar effect may be achieved by applying tensile strain to the CNT, for which BTE calculations predict superdiffusive transport~\cite{bruns_nanotube_2021}.

\subsection{Non Fourier transport in 2D materials}

Graphene and hexagonal boron nitride (h-BN) are one-atom thin rigid sheets, thus realizing in the material world the equivalent of two-dimensional lattice models. For this reason, thermal transport in graphene, and to a lesser extent in h-BN, has been intensively investigated to unravel anomalous conduction both by experiments and simulations~\cite{Balandin2011}. 
As in the case of CNTs, NEMD simulations of graphene do not usually manage to reach saturation of the thermal conductivity, due to their inherent limitation in size and time scales. Large-scale NEMD up to $10^7$ atoms showed that the thermal conductivity of graphene does not saturate for patches up to several $\mu$m long in very good agreement with experimental measurements~\cite{xu_length-dependent_2014}. In the length range considered by this work, $\kappa\propto \log(L)$, as predicted for 2D lattice models~\cite{Wang2012}, thus supporting the hypothesis of anomalous transport. 
Logarithmic divergence of $\kappa$ was observed in both NEMD and AEMD simulations on graphene with $L<2~\mu$m~\cite{ray2019heat}.
However, further extending the size of the graphene patches to $L\sim 100\ \mu$m, shows evidence of $\kappa$ saturation~\cite{barbarino_intrinsic_2015}.

\begin{figure}[h]%
\centering
\includegraphics[width=0.95\textwidth]{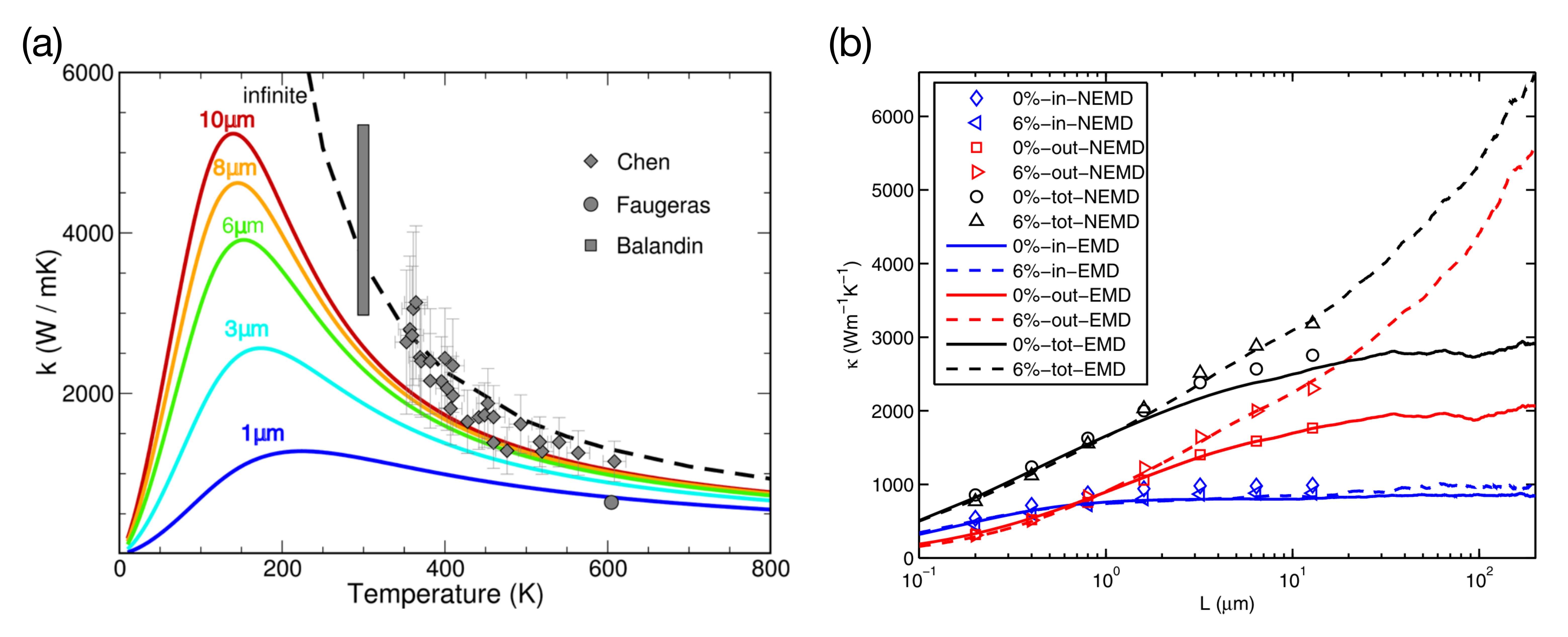}
\caption{(a) Thermal conductivity of graphene as a function of temperature and length computed solving the BTE with {\sl ab initio} interatomic force constants (adapted with permission from~\cite{fugallo_thermal_2014}. Copyright 2014, American Chemical Society). (b) Decomposition of the thermal conductivity in in-plane and out-of-plane contributions for unstrained and strained graphene from molecular dynamics simulations (adapted with permission from~\cite{fan_thermal_2017}).}
\label{fig:graphene}
\end{figure}

In fact, both EMD simulations with empirical potentials~\cite{fan_thermal_2017, fan_bimodal_2017}, and {\sl ab initio} lattice dynamics BTE calculations~\cite{fugallo_thermal_2014, cepellotti_phonon_2015, lee_hydrodynamic_2015} confirm that the extended limit of the thermal conductivity is finite. 
BTE calculations give a quantitative estimate of the bulk thermal conductivity of graphene as a function of both length and temperature in very good quantitative agreement with experiments (Figure~\ref{fig:graphene}a), and they provide in-depth insight into the transport mechanism~\cite{fugallo_thermal_2014}. 
At room temperature, heat transport is superdiffusive over lengths of the order of 1~mm, as the main heat carriers are collective phonon excitations with mean free paths that extend up to 100 $\mu$m. 
This collective effect, equivalent to Poiseuille flow in a fluid, is enhanced in low-dimensional materials, due to the reduced phase space for phonon scattering~\cite{cepellotti_phonon_2015, lee_hydrodynamic_2015}.

The decomposition of the thermal conductivity of graphene into in-plane ($\kappa_{in}$) and out-of-plane ($\kappa_{out}$) from MD simulations at room temperature shows that the relative contribution depends on the length of the graphene patch (Figure~\ref{fig:graphene}b). $\kappa_{in}$ converges at relatively short length $L\sim 1\ \mu$m, while $\kappa_{out}$ keeps increasing up to 10 $\mu$m. Remarkably, these simulations predict that strained graphene would exhibit divergent thermal conductivity, and $\kappa_{out}$, i.e. the flexural modes, would be responsible for the divergence of $\kappa(L)$~\cite{fan_thermal_2017}. This mechanism is analogous to the one discussed for strained CNT in the previous section. 
It is worth noting that some BTE calculations support the divergence of $\kappa$ in strained graphene~\cite{bonini_acoustic_2012, kuang_thermal_2016}, others suggest that strain would reduce $\kappa$, thus in stark disagreement with MD~\cite{fugallo_thermal_2014, cepellotti_phonon_2015}. The thermal conductivity of strained graphene remains an open issue.

\subsection{Ballistic-to-diffusive transport regime}

Fourier's law applies to the diffusive transport regime, which generally occurs in extended materials. However, regardless of their dimensionality, thermal transport through systems of finite size is ballistic when the mean free path of the heat carriers exceeds the distance between the heat baths. 
From measurements and simulations of nanoscale systems, an intermediate transport regime has emerged, denoted as ballistic-to-diffusive, superdiffusive, or nanoscale transport regime, in which the thermal conductivity appears to be size-dependent~\cite{majumdar_microscale_1993,hoogeboom-pot_new_2015}.
As discussed in the previous sections, in this regime the thermal conductivity depends on the length of the system along the transport direction. This length may extend to several millimeters for low-dimensional nanostructures, but superdiffusive transport has been observed over $\mu$m lengths also for bulk systems at room temperature e.g. crystalline silicon~\cite{regner_broadband_2013}, and even in the cross-plane direction of layered materials stemming from unexpectedly long heat carrier mean free paths~\cite{zhang_temperature-dependent_2016, sood_quasi-ballistic_2019}.
It is widely accepted that the superdiffusive regime emerges from the wide range of heat carriers' mean free paths, which can be accurately obtained from the exact solution of the phonon BTE at the microscopic level.\cite{cepellotti_thermal_2016}
However, it is a matter of debate whether Fourier's law can be applied to model transport at the nanoscale in ballistic-to-diffusive conditions.

Extending the work by Chen and Zeng~\cite{chen_gang_nonequilibrium_2001}, Maassen, Lundstrom, and coworkers~\cite{maassen_simple_2015, maassen_steady-state_2015, kaiser_thermal_2017} derived the Fourier's law for finite-length systems from the BTE by renouncing the assumption of local thermodynamic equilibrium, and applying physically correct boundary conditions of the forward and backward phonon fluxes at the two contacts. Assuming perfectly thermalizing boundary conditions, this approach leads to temperature jumps at the contacts, which approximates consistently the temperature profiles computed in NEMD simulations~\cite{hu_unification_2020}.  
Remarkably, one can prove that for a model system with a single mean-free-path ($\lambda$) independent on phonon frequency, the thermal conductivity over length $L<\lambda$ is the same as the bulk thermal conductivity, and
the apparent reduction of $\kappa(L)$ stems from a reduction of the net phonon flux. 
In general, this approach describes correctly both ballistic effects and the diffusive regime, thus, in principle, it can be applied at all length scales. However, this picture is an approximation, which does not take into account hydrodynamic effects. 

Hydrodynamic phonon transport effects, such as Poiseuille heat flow and second sound were observed in crystals at cryogenic conditions as early as the 1960s.
Whereas diffusive (Fourier) transport is controlled by momentum dissipating {\sl umklapp} scattering processes, hydrodynamic effects emerge from momentum conserving {\sl normal} scattering processes. The latter prevails at low temperatures, and in materials in which phonon dispersion relations limit the number of umklapp scattering processes. 
For this reason, phonon hydrodynamics was assumed to have a limited impact until recent experiments on heat transport in the nanoscale regime. 
In fact, the exact solution of the microscopic BTE, Eq.~(\ref{eq:BTE}), highlighted the contribution of collective hydrodynamic effects to heat transport in two-dimensional materials~\cite{lee_hydrodynamic_2015, cepellotti_phonon_2015}. Subsequent work showed that hydrodynamics plays a significant role not only in low-dimensional systems or at very low temperatures, but also in bulk materials at non-cryogenic temperatures~\cite{cepellotti_thermal_2016}. 
Poiseuille flow and second sound cannot be described by the macroscopic Fourier’s equation, and efforts have been made to generalize the heat equation to account for hydrodynamic effects. 

In the limit of weak crystal momentum dissipation, Guyer and Krumhansl derived a steady-state solution of the linearized BTE that can be seen as a generalization of the  Fourier's law and reads~\cite{guyer_solution_1966}:
\begin{equation}
    \vec{J} = -\kappa \nabla T + l^2\left(\nabla^2 \vec{J} + \alpha \nabla(\nabla\cdot\vec{J})\right),
    \label{eq:GKE}
\end{equation} 
with $\alpha = 2$.
This  equation is characterized by a collective non-local length $l$ that defines the length-scale $L\sim l$ at which phonon hydrodynamics impacts heat transport. 
Eq.~\ref{eq:GKE} can be obtained more generally through the Chapman-Enskog first-order expansion solution to the BTE with the relaxation time approximation~\cite{guo2015phonon}.

Similarly to $\kappa$, the hydrodynamic length $l$ can be expressed in terms of microscopic phonon properties by separating the contributions to the phonon lifetime from normal ($\tau_N$) and umklapp ($\tau_U$) scattering processes. 
In the limit of pure collective regime $l = \langle v^2\tau_N \rangle \langle \tau_U^{-1}\rangle^{-1}$~\cite{guyer_solution_1966}.
A more general expression for $l$ that incorporates also resistive processes was eventually proposed in the context of a kinetic collective model, in which the thermal conductivity is expressed as the sum of a diffusive term ($\kappa_K$) and a collective term ($\kappa_C$)~\cite{torres_first_2017, torres_emergence_2018}.

Starting from the microscopic linearized BTE, Sendra {\sl et al.}~\cite{sendra_derivation_2021} derived an expression similar to Eq.~(\ref{eq:GKE}) valid in the approximation of small but finite Knudsen number, and they provide expressions to compute the parameters $\kappa$, $l$, and $\alpha$ beyond the approximations of the kinetic collective model.
For example, in these works, the nonlocal length for bulk silicon at room temperature was estimated to be $l\sim 195$~nm.
This model produces results in accordance with the truncated L\'evy formalism with fractal dimension $<2$~\cite{vermeersch_superdiffusive_2015}, which accounts for non-Fourier transport as observed in crystalline semiconductors and semiconducting alloys~\cite{torres_emergence_2018,torres_collective_2018,beardo_hydrodynamic_2022}. 

Using the relaxons formalism, Simoncelli {\sl et al.}~\cite{simoncelli_generalization_2020} derived two coupled general equations for the temperature and the drift velocity fields, valid for heat transport from the hydrodynamic to the diffusive regime. In these equations, a thermal viscosity coefficient ($\mu$) is added to the thermal conductivity ($\kappa$) as a materials parameter. 
While the details of this formalism are beyond the scope of this review, it is notable that this work introduced a parameter that allows one to assess the deviation from Fourier's law due to phonon hydrodynamics.
This ``Fourier deviation number" (FDN) is expressed in terms of the dimensionless parameters of the newly derived viscous heat equations and correlates well with the $\mathcal{L}^2$ distance between the temperature profiles predicted by Fourier’s law and that obtained from the hydrodynamic equations.
The analysis of FDN shows that Fourier transport can emerge even at nanometer-length-scales with important ballistic contributions when hydrodynamic effects are small compared to dissipative phonon scattering, thus reconciling the viscous transport formalism with the results in Refs.~\cite{maassen_steady-state_2015, kaiser_thermal_2017}.
It is indeed likely that the application of correct boundary conditions at the thermal bath provides a qualitative Fourier-like picture of heat transport at the nanoscale with length-dependent thermal conductivity, but hydrodynamic effects are responsible in the largest part for the non-Fourier behavior observed in the simulations and experiments on low-dimensional materials and nanoscale systems, including those reported in the previous sections. 

The results of applying mesoscale hydrodynamic models to predict the thermal conductivity of nanostructures depend critically on the choice of the slip boundary conditions for the phonon flux at the surface, as for example in nanowires and thin films~\cite{guo2015phonon,sellitto2015two,guo_phonon_2018}. Atomistic MD or LD simulations can in some cases overcome this issue, as they provide a parameter-free understanding of how surfaces critically affect nanoscale heat transport, subject to the availability of reliable interatomic potentials. This has been the case for silicon-based materials, in particular nanowires and nano-membranes with rough, oxidized or nanostructured surfaces~\cite{he_microscopic_2012,duchemin_atomistic_2012,neogi_tuning_2015,xiong_native_2017,neogi_anisotropic_2020}.

\subsection{Thermal rectification}

Besides phononic effects~\cite{LiLi2012} thermal rectification may be attained by modulating electronic conduction, near-field radiation, excitonic effects in quantum structures, superconducting tunneling junctions, and reversible phase transition in phase change materials~\cite{segal_single_2008,otey_thermal_2010,basu_near-field_2011,joulain_radiative_2015,giazotto2015,ordonez-miranda_quantum_2017,wu_sufficient_2009,pekola2020,kasali_optimization_2020}.
Here we review some of the works in which the original concept of the thermal diode, based on phonons non-linearity~\cite{TerraneoCasati2002,li_thermal_2004}, is investigated in realistic (nano)materials models by atomistic simulations. 

\begin{figure}[h]%
\centering
\includegraphics[width=0.95\textwidth]{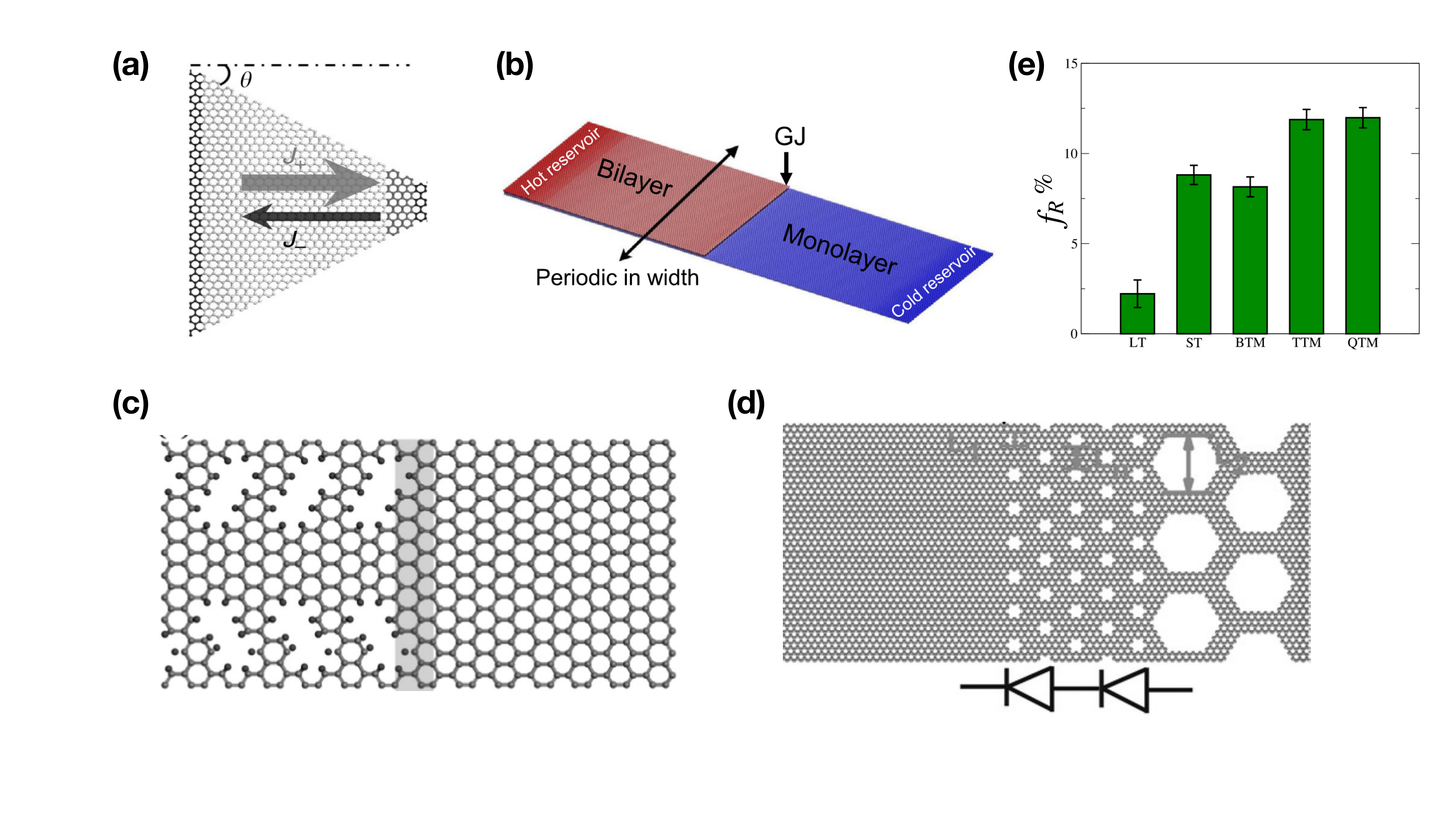}
\caption{Graphene-based thermal diodes: (a) trapezoidal patch (adapted with permission from~\cite{wang_experimental_2017}), (b) multilayer graphene step junction~\cite{zhong_thermal_2011} (adapted with permission from~\cite{rojo_thermal_2018}, Copyright 2018, IOP), (c) defect-engineered nanoribbon \cite{wang_tunable_2012} (adapted from~\cite{yousefi_thermal_2020}, Copyright 2020, Elsevier), (d) nanoporous graphene, adapted with permission from~\cite{hu_series_2017}, Copyright 2017, Wiley VCH. (e) Recalculated thermal rectification for small and large trapezoidal patches (LT, ST) and bilayer (BTM), trilayer (TTM), and quad-layer to monolayer (QTM) graphene step junctions~\cite{li_influence_2019}.}
\label{fig:graph-rect}
\end{figure}

The primary requirements for a phononic thermal diode are the asymmetry of the device with respect to the direction of the heat current and a strong temperature dependence of the phonon spectrum of the two sides of the device. 
To enable strong nonlinear effects, large temperature differences between the thermal reservoirs are necessary, which rule out any perturbative simulation method, e.g. EMD, BTE, or Landauer approaches.  
In fact, NEMD simulations have been extensively employed to calculate the rectification factor, $f_r$ in Eq.~(\ref{eq:frectification}), of several nanoscale systems and interfaces. 
Following the experimental demonstration of a thermal diode made of a mass-loaded CNT~\cite{ChangZettl2006}, carbon nanostructures have been proposed as potential thermal rectification devices. 
While NEMD simulations confirmed the relatively low rectification factors observed in experiments for mass-loaded CNTs ($f_R<16\%$)~\cite{alaghemandi_thermal_2009}, 
spectacular $f_R$ were predicted for triangular (120\% and 250\%)~\cite{hu_thermal_2009, yang_thermal_2009}, T-shaped , and trapezoidal ($\sim 100\%$)~\cite{wang_experimental_2017} graphene patches, multilayer graphene stacks (110\%)~\cite{zhong_thermal_2011}, carbon nanocones ($\gtrsim 100\%$)~\cite{yang_carbon_2008}, defect-engineered graphene nanoribbons~(80\%\cite{wang_tunable_2012}), and branched nanoribbon junctions (470\%)~\cite{chen_local_2018}.
The highest $f_R$, up to 2500\%, have been reported for CNT-graphene junctions~\cite{yang_ultrahigh_2017,yang_enhancing_2018}, modifying a design originally proposed in Ref.~\cite{alaghemandi_thermal_prb2010}.
The diode effect may be enhanced by designing interfaces in series, for example using nanoporous graphene phononic crystals, which lead to $f_R$ up to 30\%~\cite{hu_series_2017}.
Large rectification factors were also predicted by NEMD simulations of asymmetric silicon-based nanostructures, such as telescopic silicon nanowires, graded silicon-germanium alloys, and porous silicon membranes~\cite{cartoixa_thermal_2015,dettori_thermal_2016,hahn_thermal_2022}.

Recent simulations, however, suggested that the rectification factor of nanoporous or defect-engineered graphene interfaces may be much lower ($\sim 6\%$)~\cite{yousefi_non-equilibrium_2019}. Also, a joint experimental and molecular simulation study 
showed that it is unlikely to obtain detectable thermal rectification in multilayer graphene step junctions~\cite{rojo_thermal_2018}.
Further examples of discrepancies among simulations of similar systems are reported in a recent comprehensive review of simulations and experiments on thermal rectification in 2D materials~\cite{zheng_nonequilibrium_2022}.
The reasons for such discrepancies lie in the details of the simulation setups. 
The calculation of $f_r$ from NEMD is apparently straightforward, as it primarily consists of bookkeeping the heat flux, $J_{+/-}$, in the forward and backward stationary heat transport conditions as determined by the temperature of the two thermal reservoirs. 
However, the choice of the thermostats and the size of the reservoirs may lead to substantial artifacts resulting in unphysical massive thermal rectification factors. 
For example, a configuration of a graphene patch with asymmetric thermal reservoirs in NEMD simulations and the Nos\'e-Hoover (NH) thermostat \cite{hoover_canonical_1985} gives an apparent rectification factor of 920\%~\cite{jiang_remarkable_2020}. This is obviously an artifact of the simulation approach as the system is symmetric, thus violating the basic conditions to obtain a thermal diode. 
Hu{\sl et al.} proved that Langevin dynamics provides thermally equilibrated heat baths with phonon distributions consistent with the classical canonical ensemble. Conversely, with global thermostats, such as the NH and NH-chain~\cite{hoover_canonical_1985,martyna_nosehoover_1992} and velocity rescaling methods, provide a spatially uniform heat generation and nonequilibrium phonon populations in the thermal reservoirs~\cite{hu_unification_2020}.
Li {\sl et al.}~\cite{li_influence_2019} recalculated the rectification factor for trapezoid graphene patches and multilayer step junctions, which were formerly predicted to have $f_R\sim 100\%$ (Figure~\ref{fig:graph-rect}a,b)~\cite{wang_experimental_2017,zhong_thermal_2011}. The revised NEMD simulations, in which thermal baths are equilibrated with the Langevin thermostat, give much more modest rectification factors, $f_R<12\%$ (Figure~\ref{fig:graph-rect}c). 
The analysis of the phonon density of states of the thermal baths shows that using the NH thermostat promotes artificially overpopulated modes, which presence depends on the bias conditions of the system and leads to unphysically large rectification effects.
As most of the results reported above have been obtained using NH (chain) thermostats, they should be carefully ascertained using different thermalization methods or different parameters of the thermostats.  

Interestingly, a systematic reverse-NEMD study of isotope-graded CNTs reported rectification efficiencies up to 40\%, but with an opposite direction to that found in 1D graded chains: in these CNTs heat flows more efficiently from the low-mass to the high-mass end~\cite{alaghemandi_thermal_2010}. This surprising behavior is attributed to phonon scattering of longitudinal modes from perpendicular flexural modes, and it suggests that CNTs may not be used as real-world models to validate predictions for 1D systems.

\section{Experiments}
\label{sec:exp}

To conclude, we briefly mention some experimental investigations. The thermal properties of nanosized objects have an intrinsic technological interest in the field of nanoscale thermal management.  In this general context, nanowires and single-walled nanotubes have been analyzed to look for deviation from the standard Fourier's law. The typical setup to measure steady-state transport is the one sketched in Fig. \ref{fig:scheme} (see \cite{Chang2016} for a more detailed description). 

The first evidence of anomalous heat conduction in single-walled nanotubes was given in \cite{chang_breakdown_2008}. 
Further experimental evidence of anomalous transport in very long carbon nanotubes has been reported \cite{Lee2017} although the results appear controversial \cite{Li2017comment}. 
The theoretical work suggesting the importance hydrodynamic effects in room temperature heat transport in graphene and 2D materials~\cite{lee_hydrodynamic_2015,cepellotti_phonon_2015,guo_phonon_2018} fostered new experiments that lead to the observation of second sound in graphite~\cite{huberman_observation_2019}.
The wave-like nature of heat conduction, which stands at the base of hydrodynamic effects, was also proven experimentally in phononic crystal nanostructures engendering phonon interference~\cite{maire2017heat, nomura_review_2022}.
Experiments demonstrating a non-trivial length dependence of thermal conductance for molecular chains 
have also been reported \cite{Meier2014}.

Another experiment displaying anomalous transport has been reported in 
\cite{yang2021observation}. The Authors give  experimental evidence of the transition from 3D to 1D phonon transport in $Nb Se_3$ nanowires of 
less than 26 $nm$ in diameter.  
A superdiffusive thermal transport compatible with $\kappa \sim L^{1/3}$, with a large enhancement of conductivity and a normal–superdiffusive transition is observed. These changes are attributed to strong stiffening along the direction of the molecular chain direction, which induces 1D along-chain phonons dominating thermal transport. 

Other nanosystems where anomalous thermal transport is investigated are 
single-molecule junctions \cite{cui2019thermal} and trapped linear ion chains  \cite{mao2022observation}. It is conceivable that further experimental progress will occur even in this context in the next future.

\section{Conclusion}\label{sec13}
Bridging theoretical models with experiments and technological applications
is a challenging goal of physical sciences. This review aims at providing
the reader with an updated overview of the many facets, originating from the discovery that anomalous heat transport characterizes low-dimensional models of matter. Many decades ago such a scenario could appear as a mere academic curiosity, due to the absence of possible experimental verifications.
Nowadays, the exploration of states of matter at nanosize (polymers,
nanotubes, carbon layers, etc.) has opened a realistic perspective for
such investigations, together with many promising expectations of
obtaining useful recipes for designing new materials and devices.

The first part of this manuscript illustrates how non-equilibrium statistical mechanics can provide a conceptual basis for predicting the basic mechanisms of non-Fourier transport in low-dimensional materials. Beyond details of the different models studied in the past for tackling this problem, the main message is that the hydrodynamic properties of heat transport in such systems, as expected, stem from basic conservation laws. This entails universal properties, that are typical of entire classes of models, thus indicating that the search for these anomalous transport effects should correspond to fully general features, irrespective of the adopted models of matter, sharing the same symmetries.
Going through the reading of this first part one can appreciate the efforts to rationalize the overall theoretical matter in a consistent and convincing frame. In fact, the theoretical foundations of this problem presently rest on solid ground. In this respect, it is
worth pointing out that not only predictions of asymptotic, i.e
thermodynamic-limit, properties, but also finite size effects can be
suitably controlled. It can be easily realized that this is a crucial
ingredient for a reliable comparison with experiments.

The importance of anomalous transport is further stressed by its influence on thermal management and conversion. As widely discussed in the second part of this review, this unveils unexpected possibilities for thermal rectification and for enhancing beyond standard bounds the efficiency of thermodynamic motors in nanosized materials, by exploiting also coupled transport processes. 
Relying on these achievements, one can reasonably forecast the possibility of exploring very important technological applications for effective work production from thermal processes and for new energy-saving opportunities.

The last part of this review is an overview of the various computational
approaches worked out for investigating non-Fourier transport in realistic models of matter. This facet of the problem is of primary importance for checking "in silico" the signatures of anomalous transport when a direct experimental approach could be hardly effective. In fact, numerical simulations allow one to obtain reliable estimates of the finite-size effects (which typically are non-universal, i.e. model-dependent features) that may prevent the possibility of clear-cut experimental verification. 
One of the major problems is the possibility of having at disposal at least a numerical estimate of the mean-free path of thermal excitations. Only if the system under scrutiny is sensibly larger than this crucial parameter, one can reasonably expect to recover in a real sample of matter the predicted anomalous scaling properties of thermal conductivity with the system size.
A short account of a few existing experimental results is eventually outlined. The recent progress and improvements in this direction could be considered as the prelude to a rapid growth of this research field, which certainly concerns basic aspects of physical sciences, but presumably could yield a major fallout for material science and its technological applications, far from being predictable just a few
decades ago.

Finally, we point out that, while our review focused on anomalous thermal transport in classical systems, the quantum world opens up fascinating avenues.
Anomalous transport in quantum mechanical systems is much less understood than in the classical case, for both practical and fundamental reasons. 
First of all, the computational complexity in the simulation of many-body quantum systems~\cite{qcbook} grows exponentially with the number of particles, given the corresponding exponential growth of the size of the Hilbert space. 
On a more fundamental level, the same concept of local equilibrium should be carefully addressed in the quantum world. 
Indeed, we can have nanostructures smaller than the length scale over which electrons relax to a local equilibrium due to collisions with other electrons and phonons. As a consequence, effects due to quantum interference and entanglement should be taken into account in transport and energy conversion.
Moreover, strong system-reservoir coupling and memory effects, which naturally feature in small quantum systems, make problematic the same definition of heat~\cite{Hanggi2020}. Notwithstanding the difficulties, a deeper understanding of energetics in the quantum regime, and the development of ideas and tools for efficient heat management is essential for the development of quantum technologies~\cite{Auffeves2022}.

\backmatter

\bmhead{Acknowledgments}
GB, SL, DD, and RL acknowledge the contribution of the coauthors with whom they 
worked on the topics here reviewed, notably G. Casati, S. Chen, P.F. Di Cintio, S. Iubini, C. Mej{\'{\i}}a-Monasterio, A. Politi, K. Saito, J. Wang, L.F.C. Pereira, S. Neogi, S. Xiong, Z. Fan, Z. Chen, and G. Barbalinardo.

\textbf{Funding:} This work is part of the
MIUR-PRIN2017 project \textit{Coarse-grained description for non-equilibrium systems and transport phenomena} (CO-NEST)
No. 201798CZL.

\textbf{Conflict of interest:} the Authors have no relevant financial or non-financial interests to disclose.

\textbf{Authors' contributions:} All Authors contributed equally to the work. All authors read and approved the final manuscript.


\bibliography{bibnonfourier}


\end{document}